\documentclass[%
 reprint,
 amsmath,amssymb,
 aps,
prb,
]{revtex4-2}

\usepackage{graphicx}
\usepackage{dcolumn}
\usepackage{bm}
\usepackage{braket}
\usepackage{subfigure}

\begin{document}

\preprint{APS/123-QED}

\title{Dynamic nuclear spin polarization in the fractional quantum Hall effect spin transitions}
\author{Haotian Zhou}
\affiliation{Department of Physics and Astronomy, Purdue University, West Lafayette, Indiana 47907, USA}

\author{Yuli Lyanda-Geller}
\email{yuli@purdue.edu}
\affiliation{Department of Physics and Astronomy, Purdue University, West Lafayette, Indiana 47907, USA}

\date{\today}

\begin{abstract}
Experiments suggest that nuclear spins play a significant role in the quantum Hall effect (QHE) near integer and fractional QHE spin transitions, but many of these phenomena still remain to be understood. Here we study theoretically the dynamic nuclear polarization (DNP) in the two-dimensional electron liquid near a quantum point contact (QPC) or a domain wall between spin polarized and unpolarized phases induced by electrostatic gating in the fractional QHE at a filling factor 2/3 and analyze the dependence of the spin transition on temperature and the magnitude of the flowing current. We demonstrate that nearly all nuclear spins in the QPC or in the domain wall can be polarized by the electric current. The Overhauser effective magnetic field from the DNP can be strong enough to induce (or modify) a phase transition between polarized and unpolarized phases. This changes the gate voltages and magnetic fields required for the spin transitions, and leads to the reconstruction of the boundary between two phases and a domain wall and a current path displacement. The spread of nuclear spin polarization and the domain wall displacement are strongly asymmetric with respect to the direction of the current flow. Equilibration due to hyperfine interactions and its role on the nuclear spin polarization, domain wall displacements and spin transitions is studied. Back and forth oscillatory transitions between polarized and unpolarized phases near a source contact are discussed. Hyperfine interactions of nuclear spins provide a route for observation and control of the parafermion zero modes that can be induced when the domain wall between the polarized and unpolarized regions is placed in the proximity of a superconductor.
 
\end{abstract}

\maketitle

\section{Introduction}

The importance of nuclear spin systems for optical and transport phenomena in condensed matter has been long appreciated\cite{Abragam,Slichter}. Nuclear spins \cite{Rabi,Bloch,Purcell} are affected by and affect electron spins in metals \cite{Knight,Overhauser}, superconductors \cite{Hebel} and semiconductors \cite{Ekimov,Zakhar,Aronov1989,BKane,Awschalom}. One of the most intriguing settings, in which nuclear spins play a considerable role, is the fractional and integer quantum Hall effects, where coupling of electronic and nuclear spin subsystems was experimentally observed by several groups \cite{Dobers1988,Berg1990,Experiment-spin-polarized-singlet-phase,kane1992evidence,PhysRevLett.73.1011,Kronmuller,smet2002gate,Machida2002,hashimoto2002electrically,smet2004anomalous,Kou2010,Wang2021}.

The quantum Hall effect (QHE) takes place in strong magnetic fields in two-dimensional electron systems, leading to the Hall conductance quantized at $\nu e^2/2\pi\hbar$, where $e$ is the electron charge, $\hbar$ is the Planck constant, and $\nu$ is an integer or, for the fractional quantum Hall effect (FQHE) is given by certain fractions. FQHE excitations are fractionally charged quasiparticles that obey abelian 
or non-Abelian anyon statistics
\cite{lm77,wilczek1982a,wilczek1982,wilczek83,Halperin1984,Arovas1984,Read2000}. To date, the quantum Hall effect and its edge states remain one of the important settings that fuels the progress of studies of topological phenomena and quantum phase transitions \cite{Nayak2008,Lindner2012,Clarke2012,Vaezi2013,Mong2014,Vaezi2014,Csathy2018,Scarola2019,Tylan2015}. 

\begin{figure}
\centering
\includegraphics[width=0.5\textwidth]{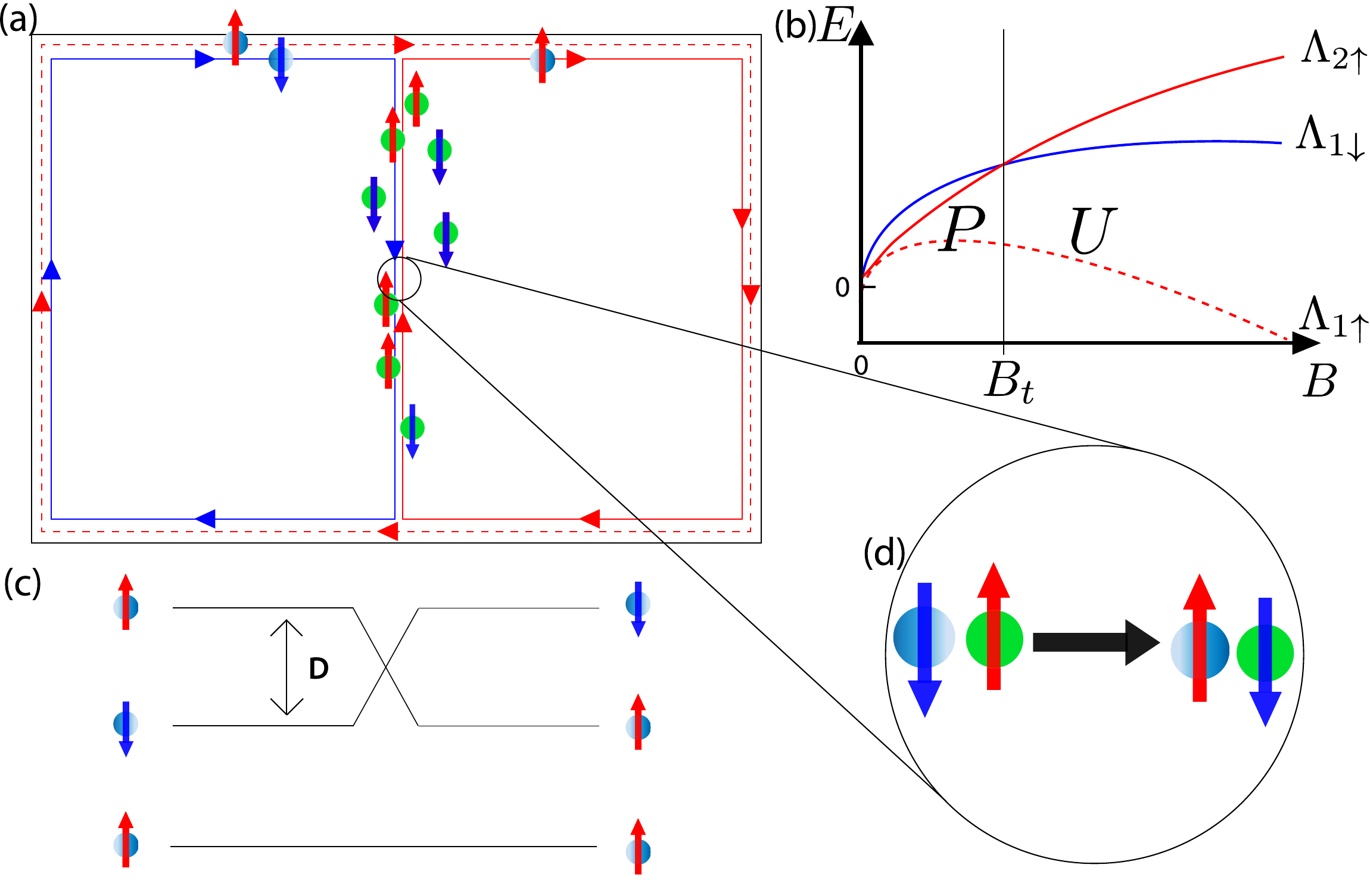}
\caption{\label{fig:illustration of the system} Quantum Hall effect spin transitions, formation of the domain wall between polarized P and unpolarized U regions, and charge carriers and nuclear spins flip-flop transitions. 
(a): The FQHE system is divided into two regions with different polarization states, due to two top electrostatic gates \cite{Wang2021}, with a domain wall separating two regions. Charge carriers are shown as blue circles, edge modes with spin up are red-colored, the edge mode with spin down is blue-colored. Besides the spin-conserving current flowing between P and U regions along the edge of the sample, additional contribution to current emerges due to mutual spin flip of nuclei (green circles) and the spin of an interacting electron in the domain wall region; 
(b) The spectrum of composite fermions in the interior of the 2D system for electron filling factor $\nu=2/3$. At the transition magnetic field $B_t$, $\Lambda_{1,\downarrow}$ level crosses $\Lambda_{2,\uparrow}$ level. The two lower fully filled states have the same spin polarization at $B<B_t$ and opposite spin polarization at $B>B_t$, resulting in polarized and unpolarized phases, correspondingly; 
(c) the energy separation between $\Lambda_{1,\downarrow}$ and $\Lambda_{2,\uparrow}$ levels in the interior of $P$ and $U$ regions, chosen to be of equal value D on two sides. Within the domain wall between two regions the separation decreases and vanishes when the order of $\Lambda_{1,\downarrow}$ and $\Lambda_{2,\uparrow}$ levels inverts, for a given $B_t$, as in panel b; (d) nuclear spins leading to correlated electron and nuclear spin flip-flop transitions.}
\end{figure}

 Recently, the quest for observation of non-Abelian statistics and non-Abelian excitations has motivated the investigation of quantum Hall systems with edge states having opposite spin polarization \cite{Clarke2012,Mong2014,Kazakov2017,Simion2018,Wu2018,Liang2019,Wang2021}. Such states naturally emerge in quantum Hall systems near spin transitions. The example of spin transition in the fractional quantum Hall liquid at a filling factor 2/3 is illustrated in Fig.~\ref{fig:illustration of the system}b, showing the spectrum of composite fermion states. The cyclotron energy of the composite fermions is defined by Coulomb interactions in the zeroth electron Landau level, $e^2/\epsilon\lambda$, where $\epsilon$ is the dielectric constant and $\lambda$ is defined by the magnetic length, which has a square root dependence on the magnetic field. The Zeeman splitting of the composite fermions is linear in magnetic fields. It is inherited from the zeroth Landau level electrons, where interactions do not lead to the Landau level mixing and associated enhancements of the spin splitting discussed in \cite{aleiner1995two,ando1974theory}.  At electron filling factor 2/3, which corresponds to filling factor 2 of composite fermions, the competition between Zeeman and cyclotron energies results in a crossing of composite fermion levels with opposite spins, so that two filled levels have either the same polarization in the polarized phase, or opposite spin polarization in the unpolarized phase.     
It was shown in \cite{Wang2021} that applying two different gate voltages to two neighboring regions of the 2D electron liquid in a triangular quantum well can lead to two neighboring domains of polarized and unpolarized quantum Hall liquids with two counter-propagating states of opposite spin polarization at the boundary between the two domains. Experiments demonstrated a crucial role of hyperfine coupling between electron and nuclear spins in this system, in which energy conservation in the flip-flops of composite fermion spins (spins of strongly-correlated electrons)  and nuclear spins is possible in the vicinity of the transition point. 

In the present paper, we develop the theory of dynamic nuclear spin polarization in fractional (and integer) quantum Hall systems. We demonstrate the Overhauser effect in tunneling between polarized and unpolarized phases of the fractional quantum Hall liquid and calculate the electric current flowing at the electrostatically induced boundary between these phases and the dynamic nuclear spin polarization of the system. In contrast to settings with spin transitions induced by variation of a global magnetic field or a uniform electrostatic gate affecting the 2D electrons, when, due to disorder, the system breaks into numerous domains of polarized and unpolarized quantum Hall liquids \cite{smet2002gate,smet2004anomalous}, in a system with two macroscopic neighboring domains we relate the nuclear spin polarization in the domain wall between the two regions to the experimentally observable value of tunneling current.
Furthermore, we show that the nuclear spin polarization that arises first at the boundary between the domains spreads next to the surrounding quantum Hall liquid because of the displacement of the boundary due to the effect of the Overhauser nuclear spin magnetic field on the fractional quantum Hall effect spin transition, as well as due to spin diffusion caused by the dipole-dipole interactions of nuclear spins. We demonstrate that the spread of nuclear spin polarization and the domain wall displacement are strongly asymmetric and occur primarily in the injector region. The displacement of the boundary between polarized and unpolarized fractional quantum Hall liquids  alters the path of the current, which explains the experimental observations in such a system \cite{Rokhinson_1}. We predict that due to displacement of the boundary between phases, in certain experimental configurations it is possible to polarize nuclear spins in almost all of the sample. We also find that if the displacement of the domain wall results in the change of polarization state of the fractional quantum Hall liquid at the source contact, the area that includes the contact undergoes an oscillatory back and forth spin transitions between polarized and unpolarized states. We discuss the role of equilibration and localization effects on the nuclear spin polarization, domain wall displacements and fractional quantum Hall effect spin transitions. We also investigate how nuclear spins provide a route for observation of the parafermion zero modes that could be induced when the domain wall between the polarized and unpolarized regions is placed in the proximity of a superconductor.
 
Despite a long history of research of nuclear spin polarization in a quantum Hall system, see e.g. \cite{Smet}, no microscopic quantum theory of non-equilibrium spin polarization due to electric current has been presented to date to the best of our knowledge, and we aim to fill this gap here. In integer quantum Hall systems, the effects of electric fields on dynamic nuclear spin polarization were qualitatively discussed in terms of the spin diode model \cite{kane1992evidence,melloch}.   Similar qualitative considerations were used to discuss the effect of a "hyperfine battery" \cite{hyperfinebattery}. Recently, several theoretical works explored the effects of dynamic nuclear spin polarization in the electron system at the edges of topological insulators \cite{Rosenow,loss-nuclear-spin-TI,Adagideli,PhysRevB.98.235412}. It has been proposed that nuclear spins lead to the enhancement of the spin orbit effects and that the effective Ruderman-Kittel-Kasuya-Yosida (RKKY) nuclear spin interaction produces a spin texture \cite{loss-nuclear-spin-TI}. It was also demonstrated that a localized magnetic moment causes backscattering  in transport through the edge states of topological insulators\cite{Vayrynen}. An obvious difference of this system with the integer and fractional quantum Hall systems is the presence of a strong magnetic field in the quantum Hall effect settings. Our main story here, however, is the consideration of the nuclear spin interactions with electron excitations of Luttinger liquids in fractional quantum Hall edge modes.

Our paper is organized as follows: in Sec.~\ref{sec:Dynamic nuclear spin polarization in a quantum point contact}, we consider dynamic nuclear spin polarization in a quantum point contact.
We derive the relation between the magnetization of current carriers and the current for integer and fractional Hall systems. For filling factor 2/3 system, we introduce a simplified approach for coupling of composite fermion and nuclear spin and treat propagating modes via K-matrix formalism. We perform a renormalization group analysis of hyperfine coupling and demonstrate that for energy scales relevant to the experiments \cite{Wang2021}, hyperfine coupling remains effective. In Sec.~\ref{sec:DNP on a domain wall}, we consider dynamic nuclear spin polarization for a long domain wall between unpolarized and polarized fractional quantum Hall liquids, tunneling current and study the equilibration of the modes propagating in the domain wall. In Sec.~\ref{sec:Effective magnetic field due to polarized nuclear}, we calculate the effective Overhauser magnetic field, demonstrate the spread of nuclear spin polarization due to the motion of the domain wall and alteration of the current path. Asymmetry between injector and collector regions in the spread of nuclear spin polarization and domain wall displacement is studied. We discuss possible patterns of spread of nuclear spin polarization in the quantum Hall system and oscillatory spin transitions between polarized and unpolarized states in situations when the region containing the source contact can switch its polarization phase. In Sec.~\ref{sec:fate of parafermion}, we discuss how hyperfine interactions of electron excitations with nuclear spins can induce parafermions in the domain wall between polarized and unpolarized fractional quantum Hall liquids proximity coupled to a superconductor. Although our calculations are restricted to a model of nuclear spin 1/2, in Sec.~\ref{sec:discussion}, we qualitatively discuss the ramifications of nuclear spins 3/2 in actual experimental settings. Our results are summarized in conclusions. A technical question of evaluation of the four-point correlators is relegated to the appendix.

\section{Dynamic nuclear spin polarization in a quantum point contact}\label{sec:Dynamic nuclear spin polarization in a quantum point contact}

\subsection{Basic consideration: uncorrelated electrons}

Our goal is to derive a steady state nuclear spin polarization and a spin-flip-flop-induced current.
In order to simplify the description of the phenomenon, we will make several assumptions.
We will consider first nuclear spin induced spin-flip transitions of the uncorrelated electrons with opposite spins. This setting arises in the integer quantum Hall effect ferromagnetic transitions, when the spectral crossing occurs between the electron Landau levels with opposite spins. This effect can  take place in systems with a large magnitude of the electron g-factor \cite{Giuliani1985,Koch1993}, in bilayer systems or wide quantum wells \cite{piazza1999first}, semiconductor systems with valley spectrum \cite{Zeitler2001}, or in heterostructures with magnetic ions \cite{Kazakov2016,Kazakov2017}.
In systems with magnetic ions, the spin flip transitions of electron states can arise both due to nuclear spin flips and magnetic ion flip-flop transitions. However, magnetic ions are almost 100\% polarized, while an equilibrium nuclear polarization at $B_z=4.5~{\rm T}$ and temperature $T=20~{\rm mK}$ is of the order of 5\%. Thus, hyperfine electron-nuclear spin scattering constitutes an effective mechanism for electron spin flips in systems with magnetic ions.

The domain wall in this section is modeled as a quantum point contact (QPC).
We restricted the region where the oppositely polarized electron states interact with nuclear spins to a QPC and assumed that the electron spins are aligned with the external magnetic field in the $z$ direction. We only consider here electrons on the left in the spin-up state $\ket{\uparrow}$ and electrons on the right in the spin-down state $\ket{\downarrow}$. For GaAs systems with negative g-factor,  positive $B_z$ represents an external magnetic field anti-parallel to the electron spin-up direction $\ket{\uparrow}$. 
The length scale characterizing the QPC is smaller than the mean free path associated with the electron-nuclei spins contact interaction. The left and right electron states are assumed to be equilibrated with the corresponding left or right chemical potential.

The Hamiltonian of the QPC is given by:
\begin{equation}\label{eq:theH}
H=H_{\text{e}}+H_{\text{hf}}+H_{\text{n}}+H_{\text{other}},
\end{equation} 
with the electron system represented by 
$H_{\text{e}}=\sum_{k,\alpha}(\xi_{k\alpha}-eV_\alpha)c_{k\alpha}^\dagger c_{k\alpha}$,
describing the quantum Hall propagating mode with the electron annihilation operator $c_{k\alpha}$ that removes an electron labeled by a quantum number $k$ (either the momentum quantum number in the Landau gauge or an angular momentum in the symmetric gauge); $\alpha=\uparrow,\downarrow$ corresponds to the spin up and spin down channels, $\xi_{k\alpha}$ is the eigenenergy of the propagating $\alpha$ mode, $V_\alpha$ is the applied voltage, $-V=V_\uparrow-V_\downarrow$ is the dc bias voltage. This voltage difference will be absorbed into the chemical potential difference of spin up and spin down electrons and define their Fermi-Dirac distribution functions $f_{\uparrow/\downarrow}(E)$. 

Next, the hyperfine contact interaction, which in this section is assumed to be isotropic,
$ H_{\text{hf}}=A_0\nu_0\sum_i\delta(\vec{r}-\vec{r}_i)\vec{I}_i\cdot\vec{s}(\vec{r})$,
where $\vec{s}(\vec{r})$ is the electron spin density operator, $\vec{I}_i$ is the spin operator of the $i^\text{th}$ nuclei, $\vec{r}_i$ is the position of the $i^\text{th}$ nuclei, $A_0$ is the hyperfine interaction constant, $\nu_0$ is the size of a unit cell. Furthermore, the nuclear Zeeman interaction with the external magnetic field is described by Hamiltonian
$H_{\text{n}}=\sum_ig_{\text{n}}\mu_{\text{n}}B_zI_{zi}$, where
$g_{\text{n}}$ is the nuclear g-factor and $\mu_\text{n}$ is the nuclear magneton.  $H_{\text{other}}$ contains terms leading to the relaxation of the nuclear spins, including nuclear magnetic dipole-dipole interaction. 
Applying the second quantization formalism to $H_{\text{hf}}$, we find
\begin{eqnarray}\label{eq:hhf2ndquantized}
H_{\text{hf}}=&&\sum_{k,k',i}A_{kk'i}(\frac{1}{2}c _{k\uparrow}^\dagger I_{-i} c_{k'\downarrow}+\frac{1}{2}c _{k'\downarrow}^\dagger I_{+i} c_{k\uparrow}\nonumber\\
&&+\frac{1}{2}[c^\dagger_{k\uparrow}c_{k\uparrow}-c _{k'\downarrow}^\dagger c _{k'\downarrow}]I_{zi}),
\end{eqnarray}
where  $A_{kk'i}=A_0\nu_0\psi^*_k(\vec{r}_i)\psi_{k'}(\vec{r}_i)$ and $I_{+i}=I_{-i}^\dagger=I_{xi}+iI_{yi}$, with $\psi_k(\vec{r})$ being the electron wave function of $k^{\text{th}}$ eigenstates. Assuming that all electrons are equally likely to interact with any nuclear spin species of the system, we simplify the interaction by taking  $\overline{A}=\overline{A_{kk'i}}$ , where $\overline{...}$ means coordinate averaging.   
 
We now calculate the steady state nuclear spin polarization $\langle M_z\rangle=\langle\sum_iI_{zi}\rangle$.
The Heisenberg equations of motion for $\langle M_z(t)\rangle$ are given by:
\begin{eqnarray}\label{eq:M_z EOM start}
\frac{d\langle M_z(t)\rangle}{dt}=&&-\frac{i}{\hbar}\left\{\langle [M_z(t), H_{\text{n}}(t)]\rangle+\langle [M_z(t), H_{\text{hf}}(t)]\rangle\right.\nonumber\\
&&\left.+\langle [M_z(t), H_{\text{other}}(t)]\rangle\right\}.
\end{eqnarray}
The first term on the RHS of Eq.~(\ref{eq:M_z EOM start}) vanishes because the nuclear Zeeman interaction polarizes nuclear spins in the $z$-direction only; the second term describes spin dynamics due to hyperfine interactions; the third term gives relaxation of nuclear spins due to nuclear spin dipole-dipole interactions, which also produces spin diffusion. The latter will be shown to be one of the factors that contribute to the spread of the Overhauser effective magnetic field acting on electrons from the domain wall to the neighboring regions of the quantum Hall liquid. Noticing that $[H_{\text{hf}},I_z+s_z]=0$, $\langle I_z+s_z\rangle$ is conserved under $U(1)_s$ symmetry, we can express the contribution of the second term of the RHS via the current flowing due to spin flip assisted tunneling between the two counter-propagating modes of the opposite spin polarization,  
$\frac{J_{\text{T}}(t)}{-e}$, where 
\begin{eqnarray}\label{eq:j_T}
\frac{J_{\text{T}}(t)}{-e}&&=-\frac{i}{\hbar}\langle [M_z(t), H_{\text{hf}}(t)]\rangle=\frac{i}{\hbar}\langle [s_z(t), H_{\text{hf}}(t)]\rangle\nonumber\\
&&=\frac{i}{2\hbar}\langle [\sum_{k}c _{k\uparrow}^\dagger(t)c_{k\uparrow}(t)-c _{k\downarrow}^\dagger(t)c_{k\downarrow}(t), H_{\text{hf}}(t)]\rangle \nonumber\\
&&=\frac{1}{2}\frac{d(N_\downarrow-N_\uparrow)}{dt},
\end{eqnarray}
where $N_\uparrow$ and $N_\downarrow$ are numbers of charge carriers with spin up and spin down, correspondingly.      
We use a simplified model, in which the current $J_{\text{T}}$ flows between the left and right electrodes due to the nuclei-induced spin flip of an electron with one spin direction propagating along the edge on the left leads to its tunneling into an electron mode with an opposite spin direction propagating along the edge on the right. This tunneling event is accompanied by nuclear spin flip corresponding to conservation of the total spin of electron and nuclei. Tunneling in the opposite direction, from the right to the left will result in opposite spin flip of a nucleus and a contribution to tunneling current in the opposite direction. If there is an asymmetry between the left and the right due to applied voltage or, as we shall see, in the presence of average non-equilibrium nuclear spin polarization if the temperature is non-zero, the net current and the net change in the nuclear spin polarization arise. The current is a consequence of spatial separation between electron modes with opposite spins. This spatial separation is not present in metals, where hyperfine interaction of electron and nuclear spins lead, e.g., to the Korringa mechanism of the nuclear spin relaxation \cite{korringa1950nuclear} or the Overhauser relaxation of electron spin \cite{overhauser1953paramagnetic} but not result in any electric current.  Using Eq.~({\ref{eq:j_T}}), we have 
\begin{equation} 
\frac{J_{\text{T}}(t)}{-e}=\frac{i}{\hbar}\overline{A}\sum_{k,k',i}c _{k\uparrow}^\dagger(t) I_{-i}(t) c_{k'\downarrow}(t)-c _{k'\downarrow}^\dagger(t) I_{+i}(t) c_{k\uparrow}(t).
\end{equation}
We use the relaxation time approximation for the third term of the RHS of Eq.~(\ref{eq:M_z EOM start}): $-\frac{i}{\hbar}\langle [M_z(t), H_{\text{other}}(t)]\rangle\approx\frac{\langle M_z\rangle_0-\langle M_z(t)\rangle}{\tau}$, where $\tau$ is the $T_1$ spin flip relaxation time due to dipole-dipole nuclear spin interactions, and $\langle M_z\rangle_0$ is the nuclear spin polarization in the thermal equilibrium described by the Brillouin function.  We note that if we identify the contribution of the second term due to hyperfine interactions of electron and nuclear spins in Eq.~(\ref{eq:M_z EOM start}) exclusively with the electric current term,  we presume that all electrons participating in electron-nuclear spins flip-flops contribute to the spin flip current upon the application of external voltage or in the presence of average non-equilibrium nuclear spin polarization. However potentially all other electron spins in the quantum Hall liquid, besides those that lead to the current, can couple to nuclear spins, and lead to nuclear spin relaxation. We will assume that these interactions define an additional contributions to spin relaxation time $\tau$. Thus, the nuclear spin polarization can be defined by the kinetic equation
\begin{equation}\label{eq:initial dM dt}
\frac{d\langle M_z(t)\rangle}{dt}= \frac{\langle J_{\text{T}}(t)\rangle}{-e}+\frac{\langle M_z\rangle_0-\langle M_z(t)\rangle}{\tau}.
\end{equation}

We now evaluate $J_{\text{T}}(t)$. It is convenient to apply a unitary transformation $U$ \cite{nazarov-glazman} to Hamiltonian in Eq.~(\ref{eq:theH}),
\begin{equation}\label{eq:unitary transform}
U=e^{-\frac{i}{\hbar}\int^tdt'E_{\text{nz}}\sum_iI_{zi}},
\end{equation}
where $E_{\text{nz}}=g_{\text{n}}\mu_{\text{n}}B_z$ is the nuclear Zeeman energy.
We first transform $H'= H-i\hbar\frac{\partial}{\partial t}$:
\begin{eqnarray}\label{eq:H'}
&&UH'U^\dagger=UHU^\dagger-i\hbar\frac{\partial U}{\partial t}U^\dagger\nonumber\\
=&&\sum_{k,\alpha}(\xi_{k\alpha}-eV_\alpha)c_{k\alpha}^\dagger c_{k\alpha}+ \sum_{k,k',i}\overline{A}(\frac{1}{2}c _{k\uparrow}^\dagger I_{-i}e^{-iE_{\text{nz}}t/\hbar} c_{k'\downarrow}\nonumber\\
&&+\frac{1}{2}c _{k'\downarrow}^\dagger I_{+i}e^{iE_{\text{nz}}t/\hbar} c_{k\uparrow}+\frac{1}{2}[c^\dagger_{k\uparrow}c_{k\uparrow}-c _{k'\downarrow}^\dagger c _{k'\downarrow}]I_{zi})\nonumber\\
&&+UH_{\text{other}}U^\dagger.
\end{eqnarray}
Here, the transform of $H_{\text{n}}$ vanishes and operators $I_{\pm i}$ gain time-dependent factors $I_{\pm i}\rightarrow I_{\pm i}e^{\pm iE_{\text{nz}}t/\hbar}$. We denote the first term of  Eq.~(\ref{eq:H'}) as $H'_\text{e}$, the second term as $H'_{\text{hf}}$ and the third term as $H'_{\text{other}}$. We then make a transform to the interaction representation with $H_0=H'_\text{e}+H'_\text{other}$, $H_\text{int}=H'_{\text{hf}}$. Then, $c_{k\alpha}(t)=c_{k\alpha}e^{-i(\xi_{k\alpha}-eV_\alpha)t/\hbar}$. 
Since the electric current arises as a result of electron tunneling caused by scattering, we apply the Kubo formula:
\begin{eqnarray}\label{eq:current_kubo_formula}
\langle J_{\text{T}}(t)\rangle=&&-\frac{e}{\hbar^2}\int^\infty_{-\infty}dt'\theta(t-t')\langle[\sum_{k,k',i}\Xi^\dagger_{kk'i}(t)-\Xi_{kk'i}(t),\nonumber\\
&&\sum_{\tilde{k},\tilde{k}',j}\Xi_{\tilde{k}\tilde{k}'j}(t')+\Xi^\dagger_{\tilde{k}\tilde{k}'j}(t')]\rangle_0,
\end{eqnarray}
where
\begin{equation}
\Xi_{kk'i}(t')=\frac{\overline{A}}{2}e^{iE_{\text{nz}}t'/\hbar}c_{k'\downarrow}^\dagger(t') I_{+i}(t') c_{k\uparrow}(t'). 
\end{equation}
We are primarily interested in a steady state case, for which the configuration of nuclear spins is nearly frozen and therefore $\vec{I}_i(t')$ in Eq.~(\ref{eq:current_kubo_formula}) is treated as $\vec{I}_i(t)$. Even in the time-dependent picture, such approximation is still valid if the evolution of the nuclear spin subsystem due to $H'_{\text{other}}$ is much slower compared to the time evolution of correlations in an electron system. Indeed, the nuclear spin relaxation time due to the dipole-dipole spin interaction is $10^{-4}s$, orders of magnitude larger than the time scale of hyperfine interaction $10^{-6}s$ \cite{JohnSchliemann_2003}; $10^{-3}s$ and longer time scales for nuclear spins may characterize quantum dots in the regime of Coulomb blockade \cite{Aleiner2002}. Hence, the nuclear spin can be treated as remaining in the same state. Furthermore, because for the QHE spin transitions problems we are primarily interested in the regime of sufficiently strong nuclear spin polarization, the nuclear spin correlations effects described by terms such as $\langle I_{+i}(t)I_{-j}(t)\rangle_0(i\neq j)$ are neglected. We also assume that the electron and nuclear correlators are separable. After these simplifications, we obtain that
$\langle J_{\text{T}}(t)\rangle$ is given by:
\begin{eqnarray}\label{eq:near final j}
\langle J_{\text{T}}(t)\rangle=&&-\frac{2e}{\hbar^2}\mathfrak{Re}\sum_{k,k',i}\int^0_{-\infty}dt'e^{i(-\xi_{k'\downarrow}-E_{\text{nz}}+\xi_{k\uparrow}+eV)(t')/\hbar}\nonumber\\
&&\times\langle[\Xi^\dagger_{kk'i}(t'),\Xi_{kk'i}(0)]\rangle,
\end{eqnarray}
where $\mathfrak{Re}$ denotes the real part of an expression.  Assuming the translation invariance of the system and using the identity $\int^0_{-\infty}dte^{-iwt}=\pi\delta(w)+i\mathcal{P}\frac{1}{w}$, with $\mathcal{P}$ being the principal value of the integral, we arrive at the following expression for the current:
\begin{eqnarray}
&&\langle J_{\text{T}}(t)\rangle=\frac{\pi\rho_e^2 e\overline{A}^2}{2\hbar}\sum_i\int d\xi\times\nonumber\\
&&\left(\langle I_{+i}(t)I_{-i}(t)\rangle_0 f_\uparrow(\xi-E_{\text{nz}})(1-f_\downarrow(\xi))\right.\nonumber\\
&&\left.-\langle I_{-i}(t)I_{+i}(t)\rangle_0 f_\downarrow(\xi)(1-f_\uparrow(\xi-E_{\text{nz}}))\right).
\end{eqnarray}
Here $\rho_e$ is the electron density of states and $f_{\uparrow/\downarrow}(E)$ is the Fermi-Dirac distribution function. Noticing that 
\begin{equation}
\label{nuclear_bias}
 \langle I_{\pm i}(t)I_{\mp i}(t)\rangle=  \langle\vec{I}_i^2-I_{zi}^2\pm I_{zi}\rangle   
\end{equation}, 
we observe that our result, which depends on the term $\sum_i\langle I_{zi}^2\rangle$ and is non-linear in nuclear spin operators, differs from that of \cite{Rosenow}.  

For nuclear spins $I>\frac{1}{2}$, this term complicates the consideration of non-equilibrium properties of nuclei. Nuclear magnetization $M_z$ describes the difference in the population of two nuclear sub-levels for $I=\frac{1}{2}$.
For three or more nuclear spin sub-levels equations are needed for two or more population differences \cite{Bloembergen1948,Dyakonov1973}, or, in other terms, Eq.(\ref{eq:initial dM dt}) needs to be supplemented by equations for $M_z^2=\langle (\sum_i I_{zi})^2\rangle$ and possibly for averages of other powers of $M_z$ for higher spins. As discussed in \cite{Bloembergen1948}, a simplified approach using nuclear spin temperature can only be used if the populations of successive levels are related by the same factor, and does not work in a general nonequilibrium case.
We note that generally consideration of full evolution of spins requires also equations for transverse components of magnetization, and correlators of powers of the components of magnetization, e.g., the simplest case known is Bloch oscillations.
However, for our purpose of investigating QHE spin transitions, equations for longitudinal component of spin polarizations are sufficient.

In what follows, we assume the nuclear spin $I=1/2$. Inserting $\langle J_{\text{T}}(t)\rangle$ back to Eq.~(\ref{eq:initial dM dt}), in the case of dipole-dipole nuclear spin relaxation time $\tau\to\infty$, the total steady state nuclear spin polarization can be written as
\begin{equation}
\label{average_spin_asymmetry} 
\langle M_z\rangle_{\text{steady}}\stackrel{\tau\to\infty}{=}\frac{N_n}{2}\tanh(\frac{\beta}{2}[eV-E_{\text{nz}}]),
\end{equation}
where $N_n$ is total number of nuclei and $\beta=1/k_BT$, $T$ is the temperature and $k_B$ is the Boltzmann constant. 
The steady state is characterized by zero tunneling current and nuclear polarization$\langle M_z\rangle$ defined by the difference in electrochemical potentials on the left and right caused by the applied voltage $V$. This occurs in much the same way as for the Overhauser effect in the case of non-equilibrium in the electron system caused by electron spin transitions between electron Zeeman levels with energy splitting $g_e\mu_e H$, when the nuclear spin polarization is defined by the electron Zeeman splitting. We note that at zero applied bias, and in the case when the nuclear spin polarization corresponds to equilibrium and is defined by the nuclear Zeeman splitting, no current flows in the system. However, if a non-equilibrium nuclear spin polarization exists in the system,
Eq.(\ref{average_spin_asymmetry}) gives an asymmetric expression for the left to the right and the right to the left transitions, and at non-zero temperature, the electric current can emerge even if the applied voltage is zero.   
For a finite $\tau$, the steady state nuclear spin polarization in the interacting region is
\begin{eqnarray}\label{eq:i=1/2 nucleus polarization}
&&\langle M_z\rangle_{\text{steady}}=\nonumber\\
&&\frac{\langle M_z\rangle_0/\tau+N_n\pi\rho_e^2 \overline{A}^2(eV-E_{\text{nz}})/4\hbar}{1/\tau+\pi\rho_e^2\overline{A}^2(eV-E_{\text{nz}})\coth(\frac{\beta}{2}[eV-E_{\text{nz}}])/2\hbar}.
\end{eqnarray}
This result is slightly different from that obtained for nuclear spin $I=1/2$ in \cite{Rosenow} due to the presence of the magnetic field in the QHE setting and the corresponding nonzero nuclear spin splitting. Notably, at zero bias, this$\langle M_z\rangle$ becomes equilibrium nuclear spin polarization:
\begin{equation}
\langle M_z\rangle_{0}=\frac{N_n}{2}\tanh(\frac{\beta}{2}[-E_{\text{nz}}]),
\end{equation}
corresponding to zero spin-flip tunneling current flowing in equilibrium.
We note that any source of generation of a non-equilibrium spin polarization, not only in the nuclear but also in the electron subsystem, and, correspondingly, extra source of generation in Eq.(\ref{eq:initial dM dt}), e.g., due to polarized light, will result in extra tunneling current at zero voltage.

It is important to discuss energy conservation in electron spin-nuclear spin transitions in the present configuration. In the interior of the 2D electron system, there is a significant mismatch between electron and nuclear Zeeman splittings, which makes energy conservation impossible. Spin-flip hyperfine transitions may only occur in the case of significant broadening of levels, e.g., when a hyperfine-induced transition is accompanied by an interaction with a thermal bath. Such events are rare at low temperatures. 

The situation is different for the edge-like modes.
Electron modes with opposite spin polarization come from different contacts subject to applied bias. The tunneling transition induced by hyperfine interactions occurs between opposite spin electron modes on the left and right. The applied bias can result in energies of tunneling electrons on the left and right different by the nuclear spin splitting.  
Indeed, the tunneling transition induced by hyperfine interaction in the quantum Hall regime is inelastic; tunneling between spin up mode and spin down mode involves decrease in energy by nuclear spin splitting, and tunneling from spin down mode to spin up mode is accompanied by an increase in energy by nuclear spin splitting. 

The nuclear spin splitting $E_{\text{nz}}$ is of the order of $10^{-7} {\rm eV}$ in the $5~{\rm T}$ magnetic field. The applied biases $V$, especially those corresponding to stronger currents needed for a sufficiently non-equilibrium regime, are much larger. Although to underscore energy conservation in the mutual electron-nuclear spin flip process we included the nuclear spin splitting $E_{\text{nz}}$  in the equation for $\langle M_z\rangle$, the corresponding energy scale is the smallest in our problem, and the nuclear spin splitting can be  omitted.

\subsection{The phase boundary in the FQHE system at $\nu= 2/3$}
It was suggested \cite{Alicea,Lindner2012,Universal_quantum} that $Z_3$ parafermions can emerge when two filling factor $1/3$ counter-propagating fractional quantum Hall modes with an opposite spin are proximitized by an s-type superconductor. The opposite spins of the modes require opposite signs of a g-factor for counterpropagating modes, which is impossible to realize in a real device. In \cite{Wu2018,Liang2019} emergence of parafermions has been suggested in systems with a filling factor $2/3$, which makes possible counterpropagating states at the boundary of polarized and unpolarized quantum Hall liquids. It was shown \cite{Wu2018,Wang2021}, that such a boundary arises when two different values of electrostatic gate voltage are applied to two neighboring regions of the $\nu=2/3$ fractional quantum Hall liquid near a fractional quantum Hall spin transition. 

The edge states of unpolarized and polarized quantum Hall liquids are theoretically described as chiral Luttinger liquids using effective theory \cite{edge_transport_Wen}.  
It was demonstrated in \cite{Liang2019} that in a sufficiently long domain wall on the boundary between polarized and unpolarized regions, Fig.~\ref{fig:illustration of the system}, away from the true edges of the sample, the counter-propagating edge-like $\nu=1/3$ states emerge, with opposite spin polarization. The approach applied in \cite{Liang2019} included an introduction of a spin vector that described the difference between polarized and unpolarized edges, analogous to the shift and orbital spin vectors introduced in \cite{edge_transport_Wen}. A more elaborate approach in \cite{Wang2021,Ponomarenko2024} showed that when charge-spin (charge-neutral) separation in chiral Luttinger liquids of the edges in the unpolarized and polarized regions and tunneling at the true edges of the sample are included, then the counter-propagating edge-like states flowing along the domain wall constitute a more complex combination of modes. 

To consider the nuclei-related effects here, we use a simple minimal model that leads to counter-propagating $\nu=1/3$ modes with opposite spins. The Abelian FQH edges are described by K-matrix theory\cite{Xiaogang,Kane}. The general action for the system is given by
\begin{equation}
S=\frac{\hbar}{4 \pi} \int dtdx \left(\mathcal{K}_{i j} \partial_t \phi_i \partial_x \phi_j-V_{i j} \partial_x \phi_i \partial_x \phi_j\right),
\end{equation}
where $i,j=1,2$ and the K-matrix of both spin singlet and polarized edge states are:
\begin{equation}
\mathcal{K}_{L / R}=\left(\begin{array}{ll}
1 & 2 \\
2 & 1
\end{array}\right).
\end{equation}
Here, $V_{ij}$ denotes the interaction between the edge states, which is a definite positive matrix. The charge density is given by 
\begin{equation}
\rho_c(x)= \frac{1}{2\pi} q_{L/R}^T \mathcal{K}_{L/R}^{-1}\partial_x\vec{\phi}=\frac{1}{2\pi}\frac{1}{3}\left(\partial_x\phi_1+\partial_x\phi_2\right),
\end{equation}
where the charge vector $q_{L/R}=\left(\begin{array}{l}1\\1\end{array}\right)$ for both polarized and unpolarized states and $\vec{\phi}_{L/R}$ are edge modes of left, right edges.
The boundary between these states is described by $\mathcal{K}=\mathcal{K}_L \oplus-\mathcal{K}_R$ and $q=q_1 \oplus q_2$:
\begin{equation}
\mathcal{K}=\left(\begin{array}{cccc}
1 & 2 & 0 & 0 \\
2 & 1 & 0 & 0 \\
0 & 0 & -1 & -2 \\
0 & 0 & -2 & -1
\end{array}\right),\quad q=\left(\begin{array}{l}
1 \\
1 \\
1 \\
1
\end{array}\right).
\end{equation}
In this combined model, we have four boson fields, which we relabel as $\phi_i(i=1,2,3,4)$. They represent the four corresponding edges $\phi_{L\uparrow},\phi_{L \downarrow}$ and two edges $\phi_{R \uparrow}$ fields of the polarized phase. 
We first consider tunneling and structure of edge-like states on the boundary of the polarized and unpolarized states in the absence of hyperfine interactions between electron and nuclear spins. In our minimal model, the difference of physical properties of polarized and unpolarized states is introduced via tunneling Hamiltonian density, which conserves spin and does not allow tunneling between edges with opposite spins:
\begin{equation}
H_T=\sum_{\vec{n}}t_{\vec{n}}(x) \psi_{\vec{n}}(x)+h.c.,
\end{equation}
where $\psi_{\vec{n}}(x)=e^{i\sum_i n_i \phi_i(x)}$, with an integer-valued vector $\vec{n}$ labeling possible configurations in the tunneling process. Since we require charge conservation, we have $q(\hat{n})=\sum_i q_i n_i=0$.
The scaling dimension of the tunneling operator is $2-\Delta(\vec{n})$\cite{Mcdonald_Haldane}, with $\Delta(\vec{n}) \geqslant \frac{1}{2}|\tilde{K}(\vec{n})|$, where $\tilde{K}(\vec{n})=\sum_{ij}n_i\mathcal{K}_{ij}n_j$. It then follows that the relevant same spin states tunneling will create an energy gap, which makes these fields massive, so that we can remove those same spin edge states from the effective low energy theory. We note that for shorter domain walls and quantum point contacts \cite{Wang2021,Ponomarenko2024}, the spin-charge separation rearranges the structure of the modes, including modes counter-propagating along the domain wall. This rearrangement is not crucial for understanding the effects of hyperfine interactions. 
For $\Delta\left(\hat{n}\right)<2$, and $K\left(n\right)=0$, we identify two sets of potentially massive fields using the criterion $\tilde{K}(\hat{n})=q(\hat{n})=0$. They are $\vec{n}_1=(0,1,0,-1)$ and $\vec{n}_2=(1,0,-1,0)$. However, because spin flip tunneling is strictly prohibited in the absence of hyperfine interactions, this set of states remains massless in low energy theory. In the next Section, we will see that from the vantage point afforded by renormalization group approach, the spin flip tunneling is irrelevant. Thus, only the same spin pair is massive and is removed from the low-energy theory. The effective theory for the two remaining fields, which we will denote $\phi_{L \downarrow}$ and $\phi_{R \uparrow}$, is characterized by the following commutation relations:
\begin{align}
[\phi_{L\downarrow}(x), \phi_{R\uparrow}(x')] &= -\frac{i\pi}{3}, \\
[\phi_{L\downarrow}(x), \phi_{L\downarrow}(x')] &= \frac{i\pi}{3} \operatorname{sgn}(x-x'), \\
[\phi_{R\uparrow}(x), \phi_{R\uparrow}(x')] &= -\frac{i\pi}{3} \operatorname{sgn}(x-x').
\end{align}
which are the commutation rules for the $\nu=\frac{1}{3}$ Hall effect combined Luttinger liquid left and right chiral fields. We therefore identify the underlying low energy effective theory as a Luttinger liquid model with the parameter $K=1/3$\cite{Giamarchi} and the following mapping:
\begin{equation}
\begin{aligned}
& \phi_{L\downarrow}(x)=K \theta(x)-\eta(x), \\
& \phi_{R \uparrow}(x)=K \theta(x)+\eta(x).
\end{aligned}
\end{equation}
Then the effective Hamiltonian at the boundary is given by
\begin{equation}\label{eq:luttinger liquid hamiltonian}
H_{\text{e}}=\frac{u\hbar}{2\pi}\int dx[K(\nabla\theta(x))^2+\frac{1}{K}(\nabla\eta(x))^2],
\end{equation}
where $u$ is edge state velocity, $u=v_c/K$. The charge edge mode velocity $v_c$ in the FQHE systems is defined by edge electronic interactions and edge potential gradients \cite{wen1991gapless,wen1990electrodynamical,Kane}. It was calculated numerically in \cite{hu2009edge}. We assume that edge modes propagating in the domain wall represent true edge modes scattered off the potential barriers formed by spatially dependent energies of the composite fermion modes, Fig.~\ref{fig:illustration of the system}(c), for respective spin states. Then edge modes propagating in the domain wall inherit the values of velocities of the true edge modes. 

We now introduce the hyperfine interactions into our system. The hyperfine interaction is due to coupling of electron spins with nuclear spins, and in our system we must define the electron charge e and spin $\vec{s}$ excitations that determine this interaction. The three-dimensional Hamiltonian of interactions of electron spin to a system of nuclear spins can be rewritten as
\begin{equation}\label{eq:hyperfine}
H_{\text{hf}}=A_0\nu_0\sum_i\vec{I}_i\cdot\vec{s}(\mathbf{r}_i),
\end{equation}
where $\vec{s}(\mathbf{r}_i)=\sum_{\alpha,\beta}\Psi^{\dagger}_{\alpha}(\mathbf{r}_i)\mathbf{\sigma}_{\alpha\beta}\Psi_{\beta}(\mathbf{r}_i)$ is an electron spin density at the site $\mathbf{r}_i$ of the nuclear spin $\mathbf{I}_i$, $\Psi_{\alpha}$, where $\alpha$ is $\downarrow$ or $\uparrow$, is the electron field operator, $\mathbf{\sigma}_{\alpha\beta}$ are the matrix elements of the Pauli spin operators. In our problem, the indices $\downarrow$ and $\uparrow$ of the electron operator correspond to the indices $\rm{L},\rm{R}$ of its longitudinal chiral modes $\Psi_{L,R}$ that correspond to oppositely moving electrons: 
\begin{equation}
\label{perp}
\Psi_{(L,R)}(\mathbf{r})=\mathit{\psi^{(L,R)}_{\perp}}(z,y)\Psi_{(L,R)}(x),
\end{equation}
where the electron operator is projected to the coordinate $x$ in which the chiral electron propagates. The coordinate $z$ characterizes an electron motion in the growth direction. The coordinate $y$ describes an in-plane direction transverse to the propagation direction of the chiral modes $x$. $\mathit{\psi^{(L,R)}_{\perp}}$ is an electron wave function in the transverse direction. Each of the longitudinal modes, in general, is characterized by its own transverse part. These modes strongly overlap in the $y$ direction and are nearly identical in the $z$ direction. In this section we will assume, similarly to \cite{loss-nuclear-spin-TI} that the electron wave function in the transverse direction is uniform (the box approximation) and
estimate that 
\begin{equation}
\label{est}
\left|\mathit{\psi_{\perp}}^{L,\dagger}(z_i,y_i)\mathit{\psi_{\perp}}^R(z_i,y_i)\right|^2=1/(\ell_z a),
\end{equation}
where $\ell_z$ and $a$ are the widths of the wave function in the $z$ and $y$ directions.
The bosonization identity is
\begin{equation}
\Psi_{(L,R)}(x)= \frac{1}{\sqrt{2\pi \delta}} \exp{\left(-i\left[K\theta(x)\pm \eta(x)\right]/K\right)},
\end{equation}
where $\delta$ is the short distance cutoff in one-dimensional continuum theory and a Klein factor is dropped. Although $\delta\simeq a$, characterizing the transverse spread of the wavefunctions of the chiral modes, we distinguish these two quantities.
Using Eqs.(\ref{eq:hyperfine}), (\ref{perp}) and (\ref{est}), and applying a standard procedure of normal ordering and point splitting, we obtain the following hyperfine interaction Hamiltonian in our one-dimensional system: 
\begin{eqnarray}
H_{\text{hf}}=&& -\frac{\nu_0}{2\pi \ell_z a}\sum_j[A_zI_{zj}\nabla\theta(x_j)\nonumber\\
&&+\frac{A_\perp}{2\delta}(I_{+j}e^{-i2\eta(x_j)/K}+I_{-j}e^{i2\eta(x_j)/K})].
\label{Hf1D}
\end{eqnarray}
We observe that in contrast to the electron-nuclear spin interactions or electron spin interactions with the magnetic impurity spin in the edge of the topological insulator, an interaction with the helical quantum Hall edge is characterized by a different imaginary exponent inverse proportional to $K=1/3$.
We will see that although the bare hyperfine constants obey $A_0=A_z=A_{\perp}$, the hyperfine interactions become anisotropic under renormalization group (RG) flow. Therefore, we have included the anisotropy of hyperfine interactions in Eq.(\ref{Hf1D}).

\subsection{FQH case: Renormalization Group analysis of the hyperfine coupling.}
We now discuss the dependence of the hyperfine coupling on energy scales, which stems from strong electron correlations of the FQH edge states modifying the behavior of hyperfine interactions of electron and nuclear spins. We will compare the results with those for the integer QHE and edge states in topological insulators.

\begin{figure}
\centering
\includegraphics[width=0.45\textwidth]{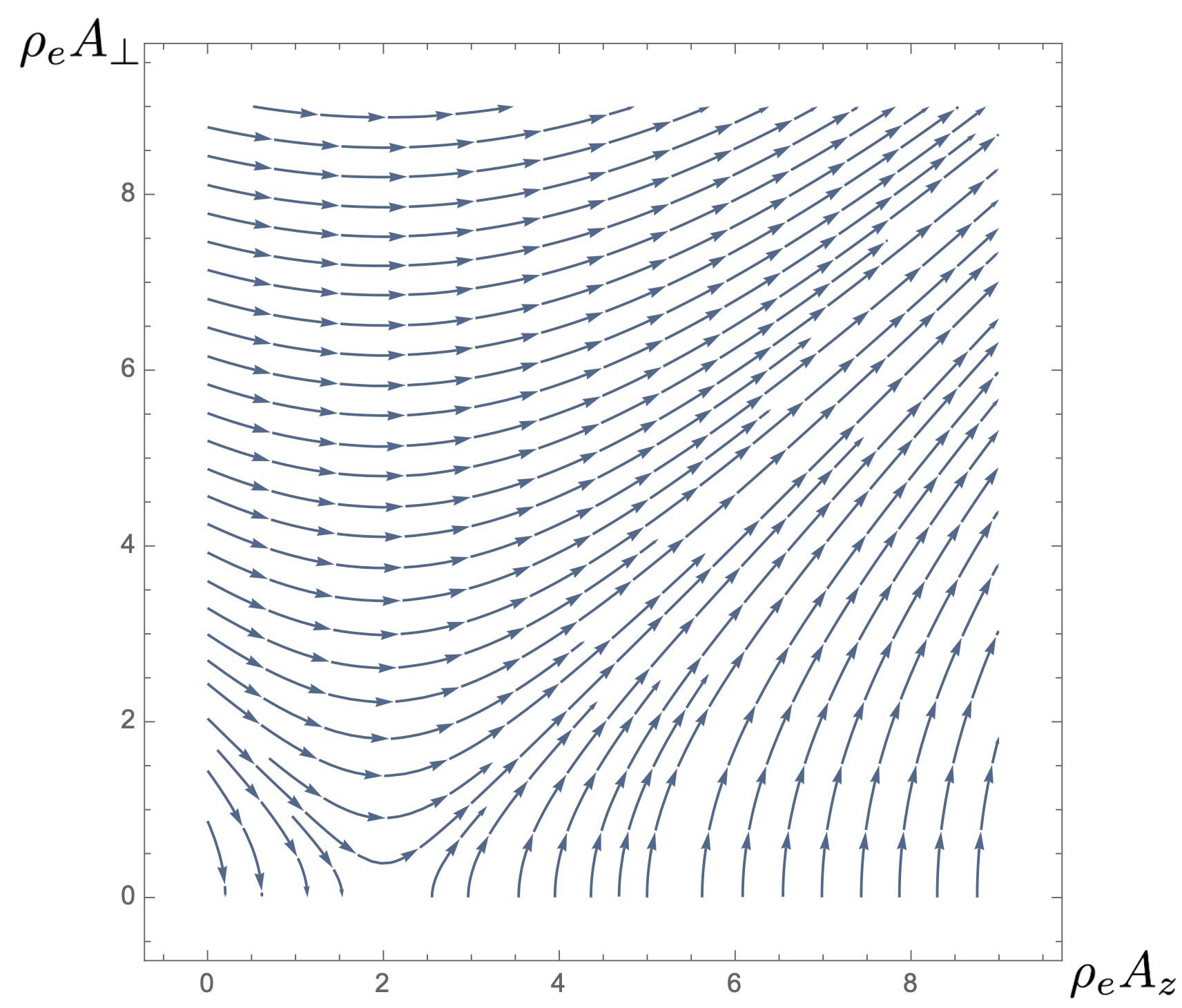}
\caption{\label{fig:RG_flow}The RG flow of hyperfine interaction between nuclear spins and electron spins for the $\nu=1/3$  FQH edge state. Only a positive part of coupling is considered, as it occurs in the GaAs system. Stable fixed points are $(0,0)$ to  $(0, 2)$ on the $\rho_eA_z$ axis, and $(\infty,\infty)$. Unstable fixed points form the line from $(0,2)$ to $(0,\infty)$ on the $\rho_eA_z$ axis. In the case of $\rho_e A<2$, when energy scale approaching $T_\text{kondo}$, the RG flows to fixed points in the $(0,2)$ region on the $\rho_eA_z$ axis, resulting in an irrelevant spin flip tunneling process. However, as discussed in the text, hyperfine coupling remains effective on energy scales that are relevant experimentally. In the case $\rho_e A>2$, the RG flows to a fixed point at $(\infty,\infty)$.   }
\end{figure}

It is important to keep in mind that the Kondo temperature estimate for the hyperfine interaction is $T_\text{Kondo}=D\exp(-\epsilon_\text{F}/A_0)$ \cite{loss-nuclear-spin-TI}, so it is unlikely that the Kondo physics associated with the hyperfine interaction can be observed in a typical range of temperatures characterizing FQHE experiments. Therefore, the perturbative methods remain valid and $\rho_e A\ll 1$ under the experimental conditions, where $\rho_e$ is the electron density of states. 
For the two opposite spin counter-propagating states at a filling factor $\nu=K$, with the edge-like states described by Eq.~(\ref{eq:luttinger liquid hamiltonian}), we arrive at the following renormalization group (RG) equations: 
\begin{align}
\frac{dA_z}{d\ln l} &= \rho_e A_\perp^2, \label{eq:fractional RG flow jz}\\
\frac{dA_\perp}{d\ln l} &= (1-1/K)A_\perp + \rho_e A_z A_\perp, \label{eq:fractional_RG_jperp}
\end{align}
The corresponding RG flow diagram is shown in Fig.~\ref{fig:RG_flow}. 
The renormalization flow is considered similarly to the case of the exchange coupling for the edge states in topological insulators\cite{PhysRevLett.96.106401,kondo_TI}. The main difference is that the tree-level diagram gives a factor $1-1/K$ rather than $1-K$ for the flow of $A_\perp$. The reason is the difference in the scaling dimensions of the electron operators in these two quasi-1D systems. For the FQH edge states, the scaling dimension is $1/K$ instead of $K$, as indicated by the exponent of the zero-temperature two-point correlator\cite{Xiaogang}. 
With our system in the perturbative regime at $\rho_e A\ll 1$, such a difference in the scaling dimensions results in a flow to a fixed point $A_\perp=0$ that corresponds to an irrelevant coupling when approaching $T_\text{Kondo}$. However, on the energy scale pertaining to the experimental conditions, the hyperfine coupling is essential. From the RG flow, we observe that the strength of the hyperfine coupling increases with higher temperature or bias voltage. This results in a stronger hyperfine interaction when an applied bias voltage results in energies of tunneling modes close to the energy of level crossing, with only two levels occupied both in the biased and grounded regions. 
Compared to the case of noninteracting electrons, e.g. in conditions of integer quantum Hall effect at $\nu=1$, in which the hyperfine interaction follows a regular Kondo RG flow, the situation is different due to the strong electron correlations in the FQH state. In the noninteracting  case, the coupling flows to a fixed point at infinity as long as $\rho_eA>0$, which means strong tunneling between edge-like states with opposite spins and effectively a disappearance of the domain wall at temperatures close to $T_\text{kondo}$. In what follows, we set the magnitude of the hyperfine coupling for our system $A_0\approx90\mu$eV\cite{Merkulov,JohnSchliemann_2003} instead of using the RG running coupling, which constitutes the average of the hyperfine coupling constants for three isotopes, $^{69}Ga$, $^{71}Ga$ $^{75}As$ in GaAs, and is the strength of $A_0$ characterizing the experimentally relevant energy scale.

\vspace{1cm}
\subsection{FQH case: Dynamic nuclear spin polarization (DNP)}
We now discuss the Overhauser effect or DNP in the FQHE case. In general, and particularly at $\nu=2/3$, the tunneling current can include contributions without and with a spin flip of a propagating charge modes. In our case, a generation of the nonequilibrium spin flip tunneling current is caused by the nuclear spin flips, leading to DNP. 
As usual for DNP, the emerging spin polarization of the nuclei is much greater than the characteristic equilibrium nuclear spin polarization.  However, the more nuclear spins are polarized, the intensity of the spin-flip transitions decreases, which leads to the suppression of the current due to spin flips. As a result of these competing tendencies, a steady state nuclear spin polarization and zero spin flip tunneling current is established at a given bias and temperature. 
This zero spin flip tunneling current corresponds to only non-spin--flip currents flowing through the system.  

We re-write the tunneling current in terms of bosonized electron operators, obtaining:
\begin{equation}\label{eq:fqhe current}
J_{\text{T}}=-2\frac{e}{\hbar^2} \mathfrak{Re}\sum_i\int^0_{-\infty} dt'\langle [\Omega^\dagger_{RLi}(t'), \Omega_{RLi}(0)]\rangle,
\end{equation}
where $\Omega_{RLi}(t')=\frac{A_0\xi}{2}e^{i(E_{\text{nz}}-eV)t'/\hbar}\psi_R^\dagger(x_i,t')I_{+ i}(t)\psi_{L}(x_i,t')$, length scale $\xi=\nu_0/\ell_za$. Inserting Eq.~(\ref{eq:fqhe current}) into Eq.~(\ref{eq:initial dM dt}), for $I=1/2$ we obtain the steady-state polarization of nuclear spins:
\begin{widetext}
\begin{equation}\label{eq:fqhe steady polarization}
\langle M_z\rangle_{\text{steady}}=\frac{\frac{\langle M_z\rangle_0}{\tau}+\frac{N_n}{2}\frac{A_0^2\xi^2}{4\hbar^2}2\mathfrak{Re}\int^0_{-\infty}dte^{i(E_{\text{nz}}-eV)t/\hbar}\langle[\tilde{\Omega}(0,t),\tilde{\Omega}^\dagger(0,0)]\rangle}{\frac{1}{\tau}+\frac{A_0^2\xi^2}{4\hbar^2}2\mathfrak{Re}\int^0_{-\infty}dte^{i(E_{\text{nz}}-eV)t/\hbar}\langle\{\tilde{\Omega}(0,t),\tilde{\Omega}^\dagger(0,0)\}\rangle},
\end{equation}
where $\tilde{\Omega}(x,t)=\psi^\dagger_L(x,t)\psi_R(x,t)$. In Appendix~\ref{ap:Evaluation of correlators}, we evaluate the correlators $ \mathfrak{Re}\int^0_{-\infty}dte^{i(E_{\text{nz}}-eV)t/\hbar}\langle[\tilde{\Omega}(0,t),\tilde{\Omega}^\dagger(0,0)]\rangle$ and $ \mathfrak{Re}\int^0_{-\infty}dte^{i(E_{\text{nz}}-eV)t/\hbar}\langle\{\tilde{\Omega}(0,t),\tilde{\Omega}^\dagger(0,0)\}\rangle$.
\end{widetext}

\begin{figure}
\centering 
\includegraphics[width=0.45\textwidth]{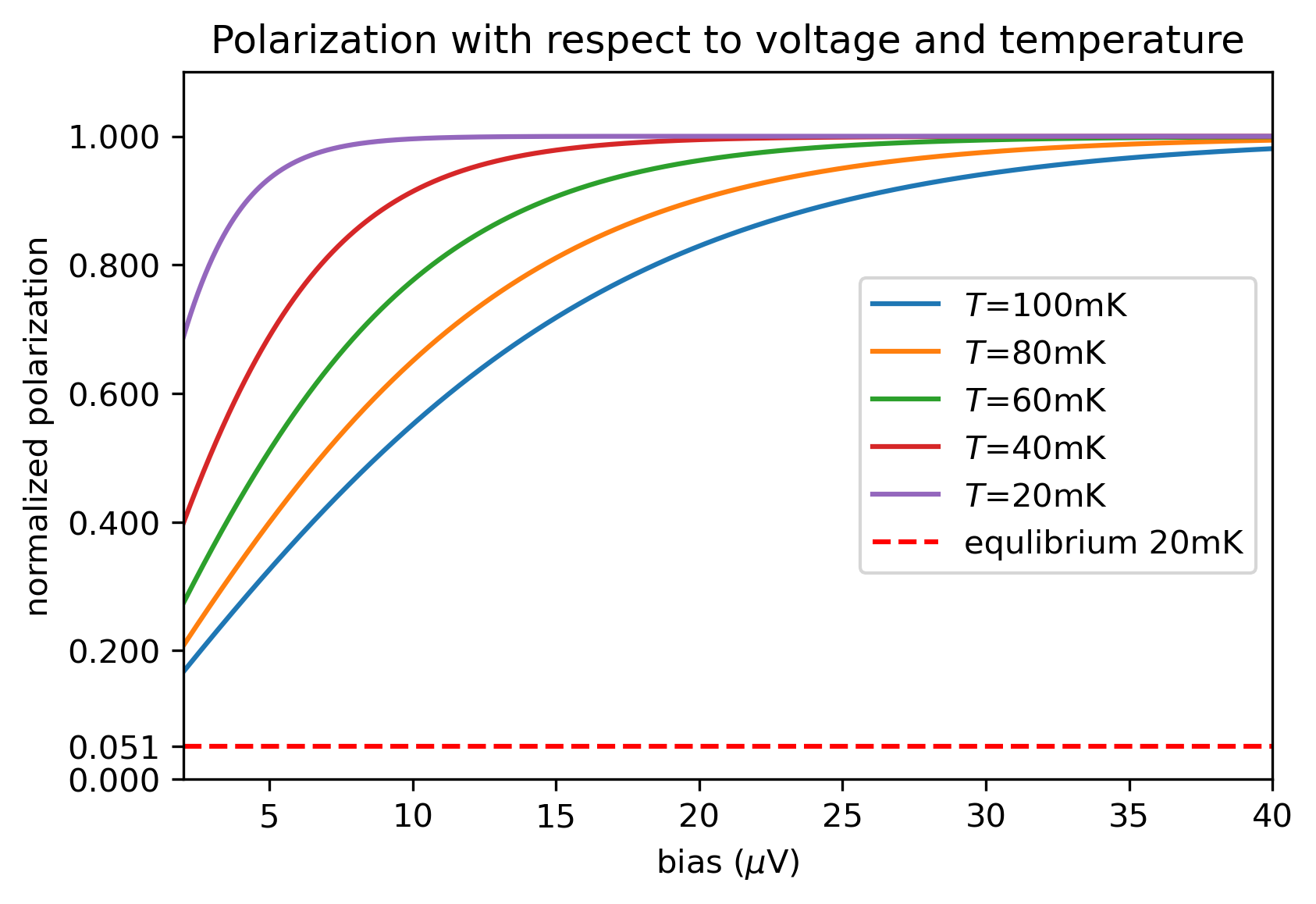}
\caption{\label{fig:polarization-voltage_dependence} The dependence of the nuclear spin polarization on temperature and bias at $1/K=3$, $B=4T$, $u=3.5*10^7$cm/s, $\tau=60$s, $\delta=10^{-6}$cm and $A_0=90\mu$eV at different temperatures in the interval between $20\text{mK}$ to $100\text{mK}$. The horizontal line is the equilibrium nuclear spin polarization under $B=4$T at $20$mK.}
\end{figure}

The evaluation of the correlators leads to the following expression for the steady-state nuclear spin polarization:
\begin{equation}\label{eq:fqhe nucleus polarization i=1/2}
\langle M_z \rangle_{\text{s}} = \frac{ \frac{\langle M_z\rangle_0}{\tau} + \frac{N_n}{2} C_1 F\left( \beta[eV-E_{\text{nz}}] \right) }{ \frac{1}{\tau} + C_1 \tilde{F}\left( \beta[eV-E_{\text{nz}}] \right) },
\end{equation}
where the constant $C_1$ is given by
\begin{equation}
C_1=\frac{A_0^2\xi^2}{8(2\pi\delta\hbar)^2} \left( \frac{\beta u \hbar}{2\pi\delta} \right)^{-2/K} \frac{\beta\hbar}{\pi},
\end{equation} 
and the functions $F(x)$ and $\tilde{F}(x)$ are 
\begin{eqnarray}
F(x) &=& B\left(1/K+\frac{ix}{2\pi}, 1/K-\frac{ix}{2\pi}\right) \sinh\left(\frac{x}{2}\right),\\
\tilde{F}(x) &=& B\left(1/K+\frac{ix}{2\pi}, 1/K-\frac{ix}{2\pi}\right) \cosh\left(\frac{x}{2}\right),
\end{eqnarray}
$B(x,y)$ is Euler Beta function. When $\tau\rightarrow\infty$, $\langle M_z\rangle_{\text{steady}}$ becomes $\frac{N_n}{2}\tanh(\frac{\beta}{2}[eV-E_{\text{nz}}])$. This is precisely the effect of the Overhauser dynamic nuclear spin polarization: non-equilibrium polarization of nuclei is essentially determined by the applied voltage rather than by the nuclear Zeeman splitting that defines equilibrium nuclear spin polarization. In order to obtain a quantitative dependence of spin polarization on voltage and temperature, we need to set the spin gaps in the interior of the polarized and unpolarized regions, the bandwidth cutoff $\delta$ and the edge velocity $u$. 
The edge state velocity $u=v_c/K$ is determined from $v_c=1.16*10^7$cm/s that follows from numerical results of \cite{hu2009edge} if to assume the value of the low-temperature dielectric constant $\epsilon=12.35$. The edge bandwidth corresponds physically to an approximate width of the edge channel, which we assume is on the order of magnetic length and therefore is approximately $10^{-6}$cm under the  magnetic field $\sim 4.5$T. Electrostatically controlled energy gaps $D$ are chosen to be equal in the polarized and unpolarized regions and smaller than or equal to the hyperfine interaction strength. We set $D=60\mu$eV. Since the relaxation of nuclear spin polarization due to nuclear spin dipole-dipole interaction in a QPC, see, e.g., \cite{Kou2010}, may take minutes, we set $\tau=60$s. We then obtain the temperature and voltage dependence of the spin polarization shown in Fig.~\ref{fig:polarization-voltage_dependence}
As expected, we can see that nuclear spins are almost fully polarized when the ratio of bias to temperature increases. 

\section{Dynamic nuclear spin polarization in a domain wall between polarized and unpolarized FQHE regions}\label{sec:DNP on a domain wall}
\subsection{The domain wall and its parameters}
We now consider the case of the domain wall between polarized and unpolarized fractional quantum Hall liquids. We consider the case of spin transition at $\nu=2/3$, but expect that our results can be applied to spin transitions at other FQHE fractions. In cases when FQHE spin transition takes place upon changing the magnetic field or applying a uniform gate to the whole 2D plane, the sample breaks into multiple domains of polarized and unpolarized quantum Hall liquids with domain walls between those. These domain walls form a percolation network responsible for the resistance peak between two plateaus corresponding to polarized and unpolarized incompressible quantum Hall liquids.  In experiments with global magnetic field or gate voltage control in \cite{Wang2021}, the peak occurs in the range of filling factors between $\nu=0.655$ and $\nu=0.67$, and two plateaus with almost zero $R_{xx}$ are formed in the ranges from $\nu=0.67$ to $\nu=0.682$ and from $\nu=0.64$ to $\nu=0.655$. When the two gates are applied to the two neighboring regions in \cite{Wang2021} in a fixed magnetic field, the two applied gate voltages correspond to the two intervals of $\nu$ above. It is reasonable to assume that the filling factors underneath the necessarily present spatial gap between two gates are in the range of filling factors, where the peak of $R_{xx}$ emerges in a set up with a global gate voltage variation. Thus, the domain wall is characterized by a gradual change of a filling factor. However, the most important feature is the character of spin states underneath two different gates, Fig.~\ref{fig:illustration of the system}. We note that from the K-matrix treatment it follows that one of the edge states propagates along the true edge of the sample, and two counter-propagating states form the domain wall with a helical channel. This naturally aligns with a picture of a helical channel flowing in a compressible region with a filling factor corresponding to the peak of  $R_{xx}$. 

The energy spectrum of the system is most easily discussed in terms of composite fermion states. In the interior of the 2D liquid, the system is characterized by an energy gap $D$ between composite fermion states of opposite spin polarization, due to electrostatic gating~\cite{Rokhinson_1}, Fig.{\ref{fig:illustration of the system}}(c). Throughout the 2D sample, $D$ is defined by the orbital and Zeeman energies of the composite fermions in the polarized and unpolarized region. We consider the GaAs-based system, in which the effective electron $g$-factor $g_e< 0$. Composite fermions inherit their Zeeman splitting from electrons, and the ground state in both regions is the spin up state. The next-lying filled state is the spin up state in the polarized region and the spin down state in the unpolarized region. The unfilled states are correspondingly the spin down state in the polarized region and the spin up state in the unpolarized regions. As the spin up and down states evolve between the unpolarized and polarized regions, they exhibit the level crossing within the domain wall, shown in Fig.~\ref{fig:illustration of the system}(c). The spin up state increasing its energy from the interior of the polarized FQHE phase to the interior of the unpolarized phase 
constitutes a barrier that defines the domain wall spin up channel, and the spin down state increasing its energy from the interior of the unpolarized FQHE phase to the interior of the polarized phase constitutes a barrier that defines the domain wall spin down channel. These channels propagate along the domain wall similarly to true edge channel propagation along the sample edge.
The edge-like states have to be filled below the energy of the level crossing and unfilled above that energy. We will assume restrictions on the applied bias that result from the conditions of the experiment \cite{Wang2021}. The experiments on nuclear spin pumping in this regime by the dc current can only use voltages that result in an electrochemical potential and energy of a tunneling mode lying below the crossing point, with only two composite fermion modes filled throughout the system. 

The domain wall differs from a QPC in that it is longer and nuclear spin induced equilibration between the edge channels with opposite spin cannot be ignored. We therefore discuss the spatial extent of a domain wall. We assume that prior to nuclear spin pumping, the domain wall strip extends from the top edge to the bottom edge of the 2D FQHE system, with a uniform width $w_D$. In the growth direction of a heterostructure, perpendicular to the 2D plane, the domain wall states are characterized by the thickness $\ell_z$ defined by the spread of the electron wavefunctions in the $z$ direction of spatial confinement. Experiments \cite{Wang2021}, investigated a triangular quantum well structure. The spread of the wavefunctions then depends on the applied electrostatic gate voltages, which differ on two sides of the domain wall. However, an estimate for an average value of $\ell_z$, about $120$\AA is sufficient for our purposes. The width of the domain wall $w_D$ is defined by the overlap of electron edge-like states. The states flowing along the domain wall are due to electrostatic gates, causing different spin polarization phases of fractional quantum Hall liquid in the two adjacent regions of 2D sample. However, the picture of edges as compressible strips that separate incompressible strips with flowing FQHE modes near the true edge of the sample discussed in \cite{Electrostatics} is not applicable here. Rather, for a very smooth change in a filling factor, as discussed above, the helical channel formed by two counter-propagating edges resides in a compressible strip. 
For an average gate potential $40~{\rm mV}$ and the electron density $n=0.9 \times 10^{12} {\rm cm^{-2}}$, we obtain the width of compressible strips for each counter-propagating mode $\ell=90$~\AA  \cite{Electrostatics}. Thus, we estimate the width of the domain wall $\sim 180$\AA. This estimate for the width also sets a scale for the inverse momentum cutoff of the Luttinger mode, $2\delta\approx w_D$, \cite{edge_transport_Wen}. We further assume that the domain wall region is characterized with a homogeneous nuclear density $\rho_{\text{n}}(\vec{x})=\rho_{\text{n}}$. This is sufficient for the qualitative picture of the region where the majority of the spin-flip scattering and tunneling occurs.  A more detailed microscopic picture of the domain wall is beyond the scope of this work.

\subsection{Tunneling current}
The tunneling current density per unit length is given by 
\begin{eqnarray}\label{eq:tunneling current}
j_{\text{T}}(x)=\rho_{\text{n}}&&\delta \ell_z(-e)C_1\nonumber\\
&&\times\{\frac{1}{2}F[\mu(x)\beta]-\langle I_z\rangle\tilde{F}[\mu(x)\beta]\}.
\end{eqnarray}
 Here we have replaced the sum over all nuclear spins with a continuum limit and $x$ refers to the position in the stripe with the origin $x=0$ at the junction of the domain wall with the top edge of the sample, where the spin flip tunneling can occur first. In Eq.~(\ref{eq:tunneling current}), we have also substituted the bias $V$ of the previous sections with a local chemical potential difference between counterpropagating channels of the domain wall $\mu(x)$. We assume that the right half of the device is grounded. Hence the chemical potential difference is just due to the chemical potential of the edge channel on the left.

 Considering the limiting case with no spin diffusion,  $\tau\to\infty$, and omitting the nuclear spin Zeeman energy, form Eq.(\ref{eq:tunneling current}) we obtain $\langle I_z\rangle=\frac{1}{2}\tanh(\frac{eV\beta}{2})$. In this case Eq.~(\ref{eq:tunneling current}) gives a vanishing steady state spin flip part of the tunneling current flowing along the domain wall. Indeed, for maximally polarized nuclear spins, spin flips can no longer generate tunneling between the modes of the opposite spin,  c.f. Eq.~(\ref{eq:i=1/2 nucleus polarization}). However, at a finite $\tau$, a residual spin flip domain wall tunneling current due to spin diffusion is possible. 
The appearance of the residual spin flip tunneling current may shed light on the experimental observations of \cite{Wang2021}. There, a nuclear spin polarization reached saturation in the presence of a high current flowing through the sample, but when a small current was applied after saturation, the observed values of the currents along the domain wall and along the true edge of the sample still do not reach the theoretically predicted \cite{Wang2021,Ponomarenko2024} values in the absence of spin flips.  The observed deviations can partially originate from the residual spin flip tunneling current due to the nuclear dipole-dipole interactions and a finite $\tau$. However, the observations can also originate from the applied small ac current not aligned with the steady-state balance of pumped nuclear spin polarization and dc current; in particular, spin-down mode on the grounded side can interact with fully polarized nuclei with spin up, tunnel and flip the spin of nuclei.

The result (\ref{eq:tunneling current}) for the spin flip induced current along the domain wall can be simplified using the following asymptotic expansion of the Gamma function:
\begin{equation}
|\Gamma(x+i y)| \approx \sqrt{2 \pi}|y|^{x-\frac{1}{2}} e^{-\pi|y| / 2}.
\end{equation}
Then for positive $\mu(x)$:
\begin{eqnarray}\label{eq:j_T expression}
j_{\text{T}}(x) \approx
e C_0\left(\frac{1}{2}-\left\langle I_z\right\rangle\right) \mu(x)^{2/K-1},
\end{eqnarray}
with the constant $C_0$ given by
\begin{equation}
C_0=\frac{ \rho_{\text{n}} A_0^2 \xi^2\ell_zw_Du^{-2/K}}{8 \pi\hbar^{2/K+1} \delta^{2-2/K}\Gamma(2/K)},
\end{equation}
where $\langle I_z\rangle=\langle M_z \rangle/N_n$, the magnetization $\langle M_z\rangle$ given by Eq.~(\ref{eq:fqhe nucleus polarization i=1/2}), defining the current density dependence on $\tau$.

The current density has a power law dependence on the bias voltage, which agrees with a general Luttinger liquid tunneling behavior~\cite{Xiaogang}. Both the spin-flip-induced and therefore the total flowing currents along the domain wall and along the true edge of the sample depend on the average polarization of nuclear spins.

\subsection{Equilibration}
As counter-propagating states flowing in the domain wall interact with nuclear spins, the local chemical potential $\mu(x)$ will change due to scattering, resulting in equilibration of electrons. Gradually, the initially unequal chemical potentials of the modes on the polarized and unpolarized sides of the domain wall, $\mu_L=eV$ and $\mu_R=0$, will equalize.  Upon tunneling from left to right along the domain wall, all charges on the right eventually flow along the true edge of the sample, towards the grounded reservoir. If the domain wall is long enough, spin-flip processes result in a substantial increase in the current along the true edge and a decrease in the current flowing along the domain wall, mimicking localization in the helical channel. If such equilibration (localization) occurs at lengths less than the width of the Hall bar (from the top to the bottom true edges of the sample), this means that all carriers from the polarized region tunnel to an unpolarized region, i.e., the domain wall essentially ceases to exist. In the experiment \cite{Wang2021}, samples with long domain walls had a diminished current along the domain wall, with almost all current flowing along the true edge of the sample.

We now consider how $\mu(x)$ changes with $x$. We take into account that the nuclear spin flip caused by the hyperfine interaction occurs on a much shorter time scale than the characteristic time of spin diffusion due to the nuclear magnetic dipole-dipole interaction \cite{JohnSchliemann_2003} and ignore the spin diffusion in this section. We will mostly consider the case of high nuclear spin polarization and assume its profile to be uniform along the domain wall at first. The equilibration process is analogous to that studied in \cite{Localization-and-conductance-in-FQH}, leading to the current dependence on $x$
\begin{equation}  
J(x)=\frac{eK}{2 \pi\hbar} \mu(x),
\end{equation}
which gives a relation between the local current and the local chemical potential of the channel. We then have: 
\begin{equation}
\frac{\partial J(x)}{\partial x}=\frac{eK}{2 \pi\hbar } \frac{\partial \mu(x)}{\partial x}=-j_{\text{T}}(x),
\end{equation}
which connects the change of $\mu(x)$ to the density of the tunneling current. Applying Eq.~(\ref{eq:j_T expression}), we obtain the chemical potential $\mu(x)$ as a function of $x$: 
\begin{equation}
\frac{\partial \mu(x)}{\partial x}=-\frac{2\pi\hbar}{K} C_0\left(\frac{1}{2}-\left\langle I_z\right\rangle\right) \mu(x)^{2/K-1}.
\end{equation}
Solving the equation, we obtain:
\begin{eqnarray}\label{eq:voltage width relation}
\mu\left(x\right)=\left\{\left[-\frac{2 \pi \hbar}{K} C_0 \right.\right.&&\left(\frac{1}{2}-\left\langle I_z\right\rangle\right)(2-2 /K)x\nonumber\\
&&\left.\left.+\mu(0)^{2-2/K}\right]\right\}^{\frac{1}{2-2/K}}.
\end{eqnarray}
This equation shows that $\mu(x)$ decreases as $x$ increases, as expected in the equilibration process. The equilibration of the domain wall states will be reached if the domain wall is sufficiently long. 

So far, we assumed that the nuclear spin polarization is uniform. However, in general, the local nuclear spin polarization depends on $\mu(x)$ and varies. The approximation of a uniform nuclear spin polarization is valid at $\mu(x) \gg 1/\beta$, because it will induce highly polarized nuclear spins. But, since $\mu(x)$ is decreasing with increasing $x$, for a sufficiently long $\ell_D$ we can reach $\mu(x)\approx 1/\beta$, so the approximation is formally no longer valid. However, because of nuclear spin diffusion, nuclear spins can possibly remain sufficiently polarized. We note that the approximation of uniform nuclear spin polarization is also valid for the case of weak currents $\mu(x)\ll \beta$, 
because an average nuclear polarization is then given by $\left\langle I_z\right\rangle \approx \frac{1}{2}\tanh(-\frac{\beta}{2}E_\text{nz})$, which describes the polarization of nuclear spins in equilibrium. In this case, spin diffusion affects the equilibration of electron excitations, which then depends on relaxation processes in the nuclear spin subsystem described by $\tau$, via Eqs.~(\ref{eq:j_T expression}) and (\ref{eq:fqhe nucleus polarization i=1/2}).

\section{Effective magnetic field due to polarized nuclear spins and its effect on the system}\label{sec:Effective magnetic field due to polarized nuclear}
The nuclear spin polarization induced by the tunneling current leads to an effective magnetic field acting on the electrons. 
This effective Overhauser magnetic field originates from the term $ s_z I_z$ in the hyperfine interaction. Fully polarized nuclear spins generate a field of around 5T in GaAs \cite{melloch, Slichter1990} that exceeds the Zeeman field caused by an external magnetic field.
In the configuration with the polarized and unpolarized regions separated by a domain wall, the dynamic polarization of the nuclear spins affects the FQHE phases, spin transition, and domain wall via the Overhauser field. 

\subsection{Injector-Collector Asymmetry of the Overhauser effect on the spin transition}\label{sub: InjectorCollectorAsymmetry}

We now show that the effect of the Overhauser field on the domain wall depends on the direction of current propagation through the system. We attribute the terms source and drain to the contacts to the fractional quantum Hall system and describe the regions of the sample adjacent to a domain wall as an injector region on the source side and a collector region on the drain side.
  
  If the current flows from the unpolarized region to the polarized region, the electron excitation flips its spin from down to up, while the nuclear spin flips from  up to down, as in Fig.4c. As the hyperfine constant $A_0$ is positive in the GaAs system, down nuclear spin will result in the nuclei-induced decrease in an electron (and composite mode) Zeeman energy. For a system with a negative g-factor such a decrease means a positive effective Zeeman magnetic field acting on electron spins. 
  Therefore, at $g_e< 0$, in the polarized region, the energies of both filled states with spin up will decrease, while the energy of the unfilled spin down state will increase, Fig.4a,b. The gap between the filled state and the unfilled state will grow, so that the polarized region moves further away from the fractional quantum Hall effect spin transition. In contrast, in the unpolarized region, while the energy of the lowest spin up state will go down, the second filled spin down state will increase its energy, and the gap between this state and the unfilled spin up state will decrease. Furthermore, it is possible that the effective nuclear spin field will result in a change of the sign of the gap and an inversion of the order of the filled and unfilled states, changing the state of the region where the nuclear field emerges from unpolarized to polarized. In the evolution of composite fermion spin separation in Fig.4a,b, the sign of the gap changes to the left of the initial centerline of the domain wall, and up to the displaced crossing between levels.  

  If the current flows from polarized to unpolarized region, the correlated electron (and composite fermion) spin flips from up to down, while the nuclear spin flips from down to up. Then, the nuclear spins generate a negative effective magnetic field acting on charge carrier spins. The energies of the filled spin up composite fermion states in the polarized phase will increase ($g_e< 0$), while the unfilled spin down state will experience the hyperfine-induced downward shift. Therefore, the gap will decrease and the inversion of the order of filled and unfilled levels will become possible. In the unpolarized region, the second filled spin down level shifts downward, while an unfilled spin up level goes up in energy. Here, the unpolarized region moves further away from the spin transition. Thus, we observe the following asymmetry: the injector region moves closer to spin transition or even can change the spin order, while the collector region moves further away from the spin transition. This asymmetry is independent of whether injector and collector region are polarized or unpolarized.

  \subsection{Phase reversal due to Overhauser field}
  
   At some combination of bias voltage and $l_D$, the order of filled modes can be inverted in the entire region of the domain wall on the injector side. This effect requires a motion of the domain wall in the transverse direction $y$. Within the limits of the domain wall model Eq.~(\ref{est}) we have used so far, we can observe how the injector side of the domain wall can change the order of filled spin modes, Fig.4a. In this model, the propagating modes are located entirely in the fixed strip of the domain wall, and nuclear spins flip within the domain wall strip as well, as in Fig.4c.
   
    For reversal of the order of filled spin states of the composite fermion modes, the induced spin splitting due to the Overhauser effect has to match the gap $\bar{D}$ between the modes and therefore the average nuclei polarization has to reach the value 
\begin{equation}\label{eq:relate polarization to that of 5T}
\frac{\bar{D}}{E_\text{\rm O}}=\tanh \left(\frac{\beta}{2} e V\right),
\end{equation}
where $E_\text{\rm O}=A_0  I $ is the magnitude of the effective Zeeman coupling due to the average hyperfine field, $I$ is the spin of the nuclei.
For GaAs systems, $ I=3/2$ and the effective Zeeman magnetic field  $B_{\rm O}=E_\text{\rm O}/g\mu_B=-5.3 T$, where $g=-0.44$ is the single-particle g-factor in the bulk GaAs crystals and wide quantum wells and $\mu_B$ is the electron Bohr magneton. Although we calculate our results using a simple model with spin 1/2, $E_\text{O}$ with $I=3/2$ is the energy scale that describes the application of this model to experimentally investigated GaAs systems.
The voltage $V$ defines the energy of the tunneling quasiparticles, which in turn defines the gap between the filled and unfilled levels $\bar{D}$. We assume for simplicity that in the absence of nuclear spin polarization the second spin up filled level in the interior of the polarized region and the spin down filled level in the interior of the unpolarized region have the same energy, and the energy separation $D$ between the second filled levels and the next unfilled levels is the same in the interior of both regions. Then the range of voltages and quasiparticle energies $\epsilon$ restricted by the limitation of two filled composite fermion levels throughout the system is $0\le eV,\epsilon\le D/2$, where we omit Zeeman nuclear spin splitting, and $D/2$ is the energy at the crossing of the filled spin up and down level in this  model. The spin splitting $\bar{D}$ between the filled and unfilled levels in the injector ranges from $D$ on the boundary of the domain wall with the interior region of the 2D quantum Hall liquid to zero at the level's crossing. Then, if $D$ is smaller than $E_\text{O}$, as soon as $eV$ exceeds $\mu_D=\frac{2}{\beta}\tanh^{-1}{\left(\frac{D}{E_O}\right)}$, the relation (\ref{eq:relate polarization to that of 5T}) is fulfilled and the side of the domain wall adjacent to the injector region with polarized nuclei will reverse its order of filled and unfilled levels and experience the transition to the same polarization state as the collector region. If $\bar{D}=D$, as it takes place at the boundary between the domain wall and the injector region in the absence of nonequilibrium spin polarization, then for fully spin polarized nuclei Eq.(\ref{eq:relate polarization to that of 5T}) and uniform nuclear spin polarization following from Eq.~(\ref{est}), energies of the spin up level and spin down level at this boundary will become degenerate, as shown in Fig. 4a.

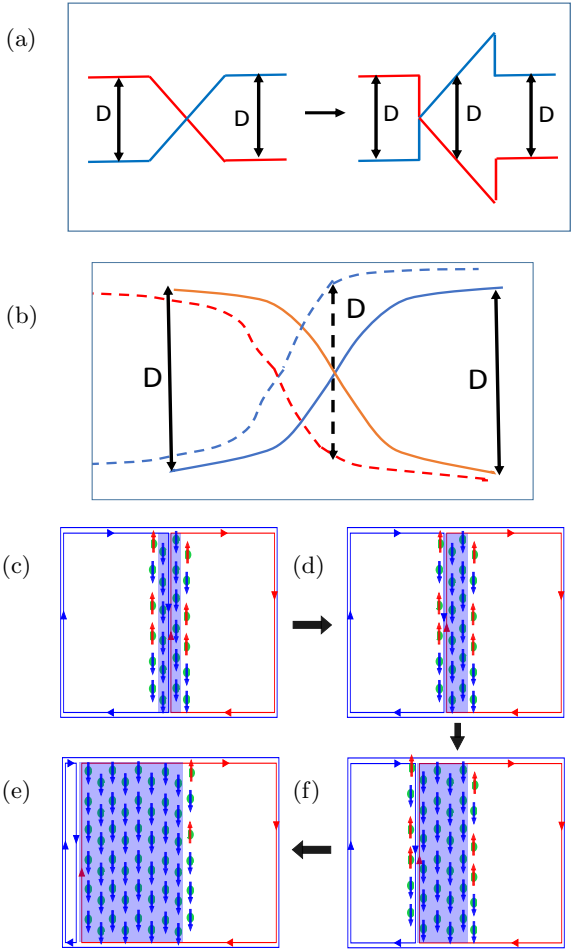
\begin{figure}
\centering
\vspace{-0.7cm}
\subfigure
{
\begin{minipage}[b]{0.46\textwidth}
\includegraphics[width=7.2cm,height=3.7cm]{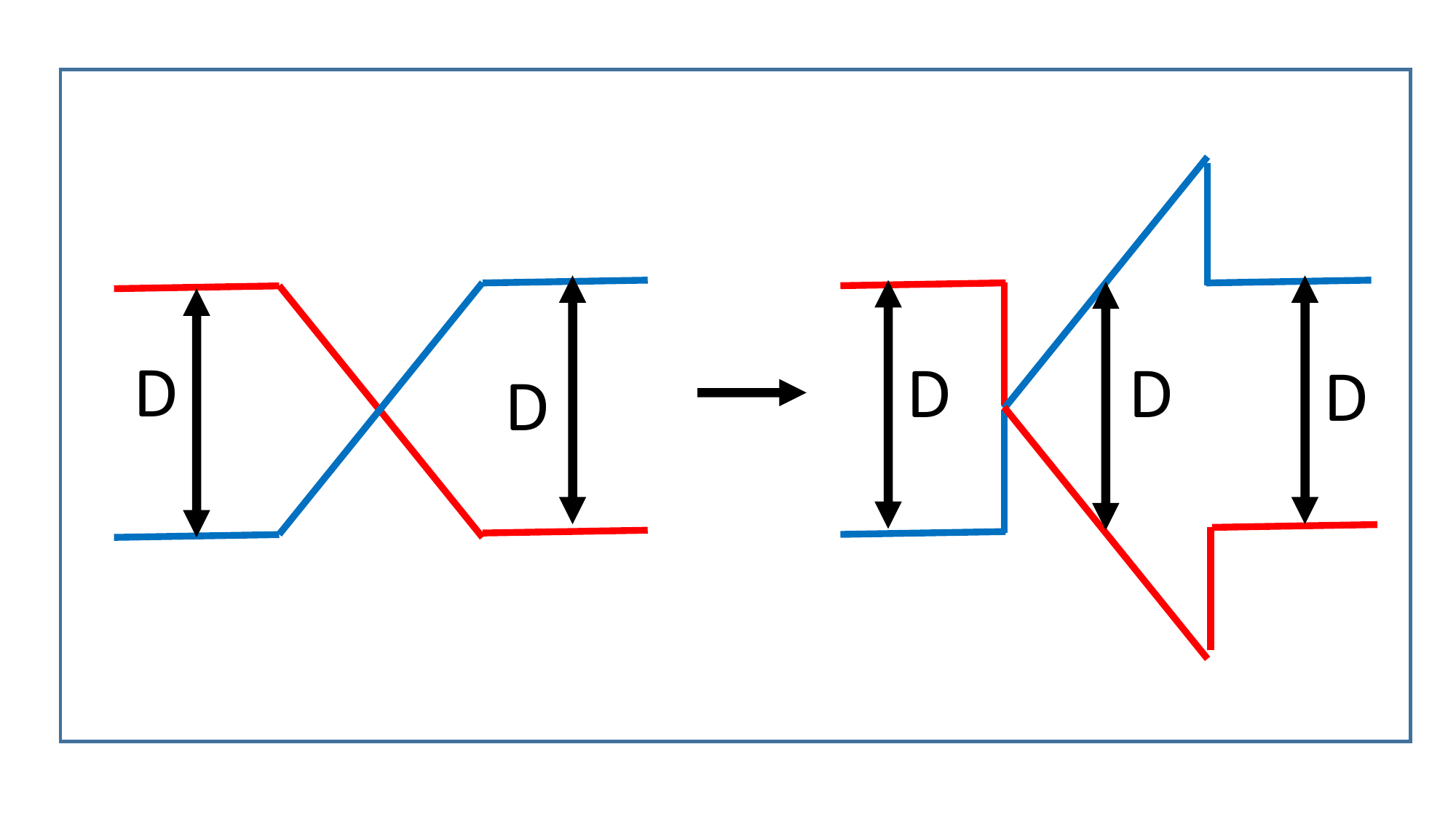}
\put(-220,80){(a)}
\end{minipage}
\label{4a}
}
\\
\vspace{-0.4cm}
\subfigure
{
\begin{minipage}[b]{0.46\textwidth}
\includegraphics[width=7.2cm,height=3.7cm]{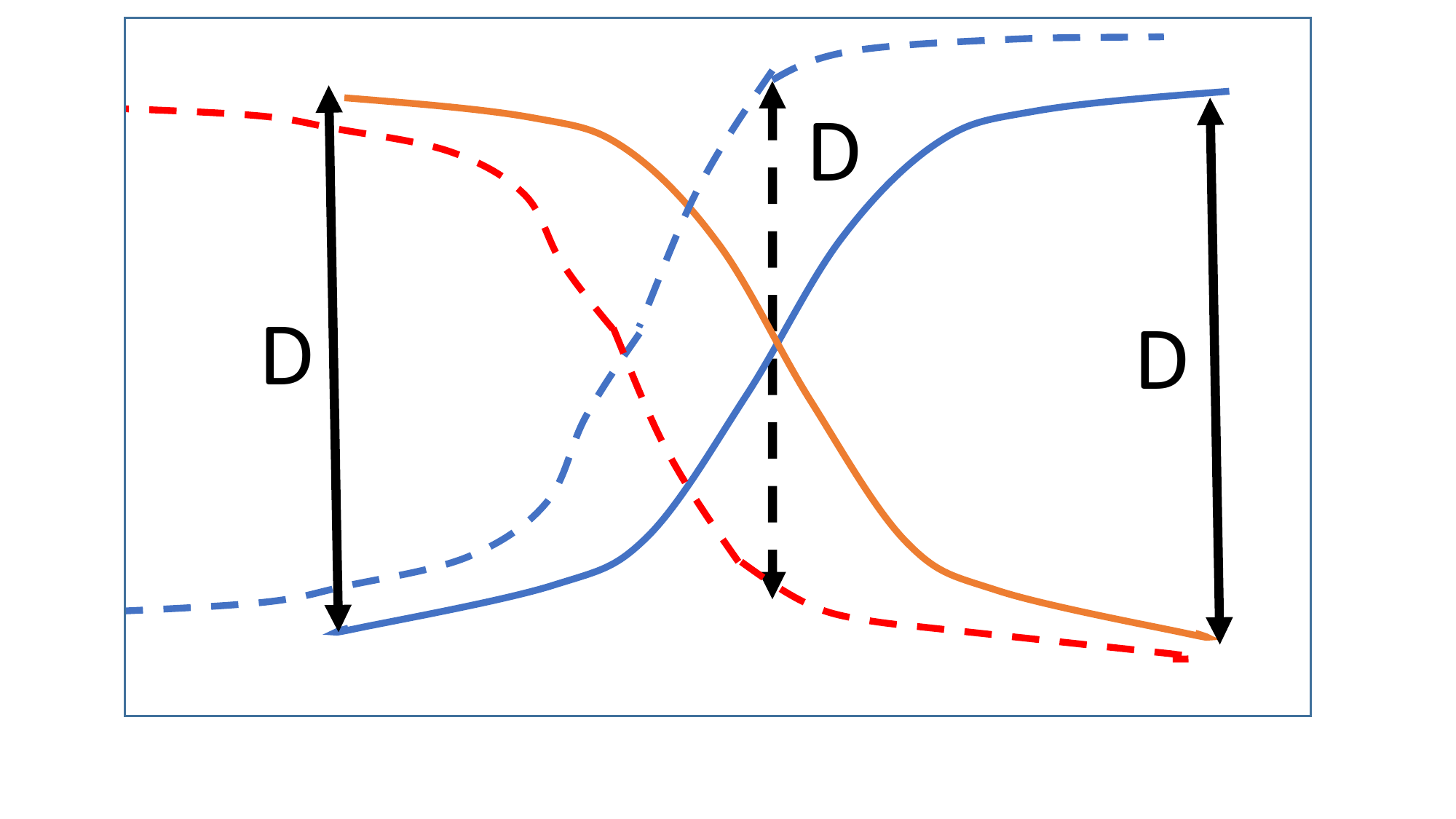}
\put(-220,80){(b)}
\end{minipage}
\label{4b}
}
\\
\vspace{-0.8cm}
\subfigure
{
\begin{minipage}[b]{0.46\textwidth}
\includegraphics[width=7.2cm,height=6.2cm]{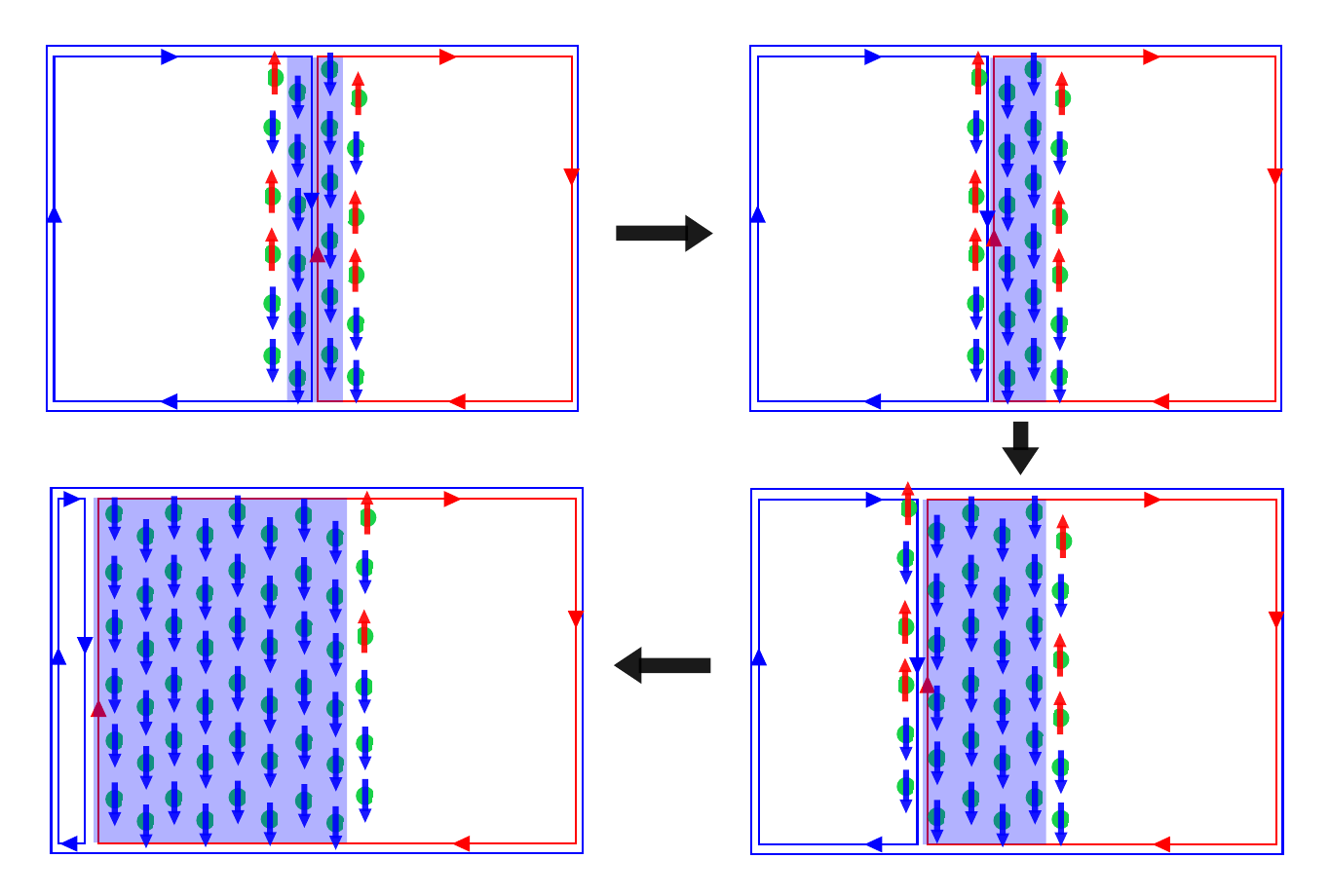}
\put(-220,150){(c)}
\put(-110,150){(d)}
\put(-220,65){(e)}
\put(-110,65){(f)}
\end{minipage}
\label{4c}
}
\vspace{-0.4cm}
\caption{\label{fig:shifting boundary} The propagation of dynamic nuclear spin polarization and a domain wall shift.
(a) (right panel) The displacement of the level crossing due to current-induced nuclear spin polarization  within the domain wall towards the boundary of the original domain wall with unpolarized nuclei. In the domain wall with unpolarized nuclei (left panel), energy separation $D$ between filled and unfilled composite fermion level is assumed equal. Current-induced spin polarization produces energy separation $D$ within the domain wall, compensating the separation in the part of the domain wall adjacent to the injector, increasing  the separation in the part adjacent to collector and leading to inverted order of filled levels at the original crossing point compared to the left half of the device. Box approximation is assumed; no nuclei are polarized outside the domain wall strip;
(b) The displacement of the domain wall beyond the box approximation. Solid lines show crossing levels in the domain walls with unpolarized nuclei. Current-induced nuclear spin polarization produces maximal spin splitting of levels at the original level crossing. Due to the Overhauser field leading to decrease in energy separation on the injector side, the domain wall is displaced, shown by displaced dashed lines for composite fermion levels;  
(c,d,e,f) Stages of the domain wall displacement. (c) Nuclear spins polarized in the domain wall; (d)  The current path follows the shift of the level crossing and shifts towards the boundary of the initial domain wall;(f) Nuclear spins are polarized in the adjacent area of the injector (beyond the box approximation) reducing energy separation between levels or inverting order of filled levels due to the Overhauser field. (e) the domain wall moves further towards the left edge of the injector region.    
}
\end{figure}

For the purpose of considering the displacement of the domain wall, we will abandon the approximation of a uniform box for the transverse part of the edge channel field operator Eq.(\ref{est}), and assume that the transverse $y-$ parts of the field operators describing the helical channels gradually decrease from the domain wall into interior regions. For helical channels at the phase boundary between the two different polarization phases in the integer quantum Hall effect, such behavior was derived in \cite{Simion2018}. In such a picture,  the passage of the current will generate non-equilibrium nuclear spin polarization in the entire range of this transverse variation. 

The domain wall can now be visualized as follows. In the spatial range of the 2D liquid beneath the spatial gap formed between two electrostatic gates on the surface of the sample, the filling factor slowly varies. Experimentally, this continuous variation corresponds to the interval of filling factors in which the current peak characterizing the fractional quantum Hall spin transition arises when measured via the variation of the electrostatic gate voltage \cite{Wang2021}. It is reasonable to assume that a spatial distribution of the current in the domain wall can be inferred from the current dependence on the filling factor that is gradually changing with transverse coordinate in the domain wall. The same spatial distribution will then describe the spin polarization of the nuclei caused by the current. We note that in a more detailed consideration of tunneling at $\nu=2/3$ taking into account spin-charge and spin-neutral mode separation \cite{Ponomarenko2024}, not all of the current flowing along the domain wall is a spin-flip current. Still, in a picture of domain wall modes gradually decaying towards interior regions, the spin-flip current has the same profile as the whole domain wall current. Then the transverse profile of the current-induced nuclear spin polarization can be described as
\begin{equation}
\label{P_N}
P_N =P_0 \exp{\left(-\frac{2y^2}{\lambda^2}\right)},
\end{equation}
where the transverse coordinate $y=0$ corresponds to the spin level crossing at the centerline of the domain wall, and $\lambda$ is the magnetic length. Compared to the integer quantum Hall spin transition case \cite{Simion2018}, we have omitted a slight asymmetry, i.e. the term linear in $y$ in the exponent, which depends on the momentum $p_x$ along the domain wall helical channel.    

We note that in the leading approximation, we consider the time scale $\tau_h$ of hyperfine interaction effects to be much shorter than the time scale of spin diffusion and nuclear spin dipole-dipole interactions. In this picture, the non-equilibrium nuclear spin polarization emerges due to the passage of the electric current on the time scale of $\tau_h\sim W^{-1}_{L\sigma; R\sigma^{\prime}}$, where $W_{L\sigma; R\sigma^{\prime}}$ is the probability of the mutual nuclear spin--spin of electron excitation flip. With the emergence of nuclear spin polarization, an average Overhauser field emerges on the same time scale, and the potential profile defining the transverse extent of the helical modes modifies. 

We now set the condition that the maximal spin polarization in the domain wall at $y=0$ is determined by Eq.~(\ref{eq:relate polarization to that of 5T}) at $\bar{D}=D$, where $D$ remains the 
energy separation between the last filled and first unfilled composite fermion state in the interior of the injector region in the modified domain wall picture. This nuclear spin polarization leads to a composite fermion energy splitting that would compensate the energy splitting in the interior of the injector phase. At this stage, however, no sufficient nuclear polarization arises in the bulk of the interior of the injecting side. Instead, this nuclear spin polarization induces an energy separation at the domain wall centerline corresponding to energy separation in the interior of the collector region, and leads to gradual compensation of the energy separation to the left of the domain wall centerline towards the injecting region, Fig. 4b. At certain negative $y=\bar{w}$, the Overhauser effective field results in energy degeneracy of the composite fermion levels, corresponding to spin transition, making  $y=\bar{w}$ a centerline of a shifted domain wall. In the interval between $y=0$ and $y=\bar{w}$, the energy separation between levels decreases, while the order of the filled levels is the same as in the collector region. 
At $y<\bar{w}$, the order of the filled levels is the same as in the injector region, and the energy separation gradually increases towards its interior. The spatial profile of the energy separation between the composite fermion levels shifts towards the source due to the Overhauser field defined by the nuclear polarization (\ref{P_N}).

Thus, the boundary between the polarized and unpolarized regions is displaced to a region where the polarization of nuclear spins is far from maximal. The energy splitting induced by the position-dependent nuclear spin polarization (\ref{P_N}) combines with the potential profile that existed for initially unpolarized nuclei to create a potential profile for edge-like modes in a shifted domain wall, Fig.4b. A new displaced current path emerges along the modified domain wall. The process is then repeated because of the new spatial profile of the polarization of nuclear spins generated along this new path. In experiments \cite{Wang2021}, measurements of currents flowing along the domain walls between contacts with a fixed position will then require a larger adjustment of the static gate on the injector side, preventing the change of the current path. This may explain the asymmetric propagation of nuclear spin polarization~\cite{Rokhinson_1,Zhong2018}. 

We now estimate the speed $v_D$ characterizing the displacement of the domain wall. The injected current $j=1{\rm nA}$ that corresponds to $\mu\approx 40\mu{\rm eV}$ leads the spin flip current $j_{sf}\sim 0.25{\rm nA}$, which results in the rate of change of the nuclear spin polarization up to $dM_z/dt= 1.5 \times 10^9 {\rm s}^{-1}$. These values correspond to Eqs.~(\ref{eq:initial dM dt}, \ref{eq:j_T expression}) and are consistent with experimental conditions for spin pumping in \cite{Wang2021}. Then the saturation of the nuclear spin polarization in the domain wall with dimensions defined by $w_D=180$\AA, $\ell_z=120$\AA, and $\ell=10\mu{\rm m}$ will be reached in $t= 0.1{\rm s}$, which will result in the domain wall displacement by $\sim 100$\AA. Thus, the speed could be up to $v_D=10^{-5} {\rm cm/s}$.  

\subsection{Role of equilibration}

\begin{figure}
\centering
\includegraphics[width=0.5\textwidth]{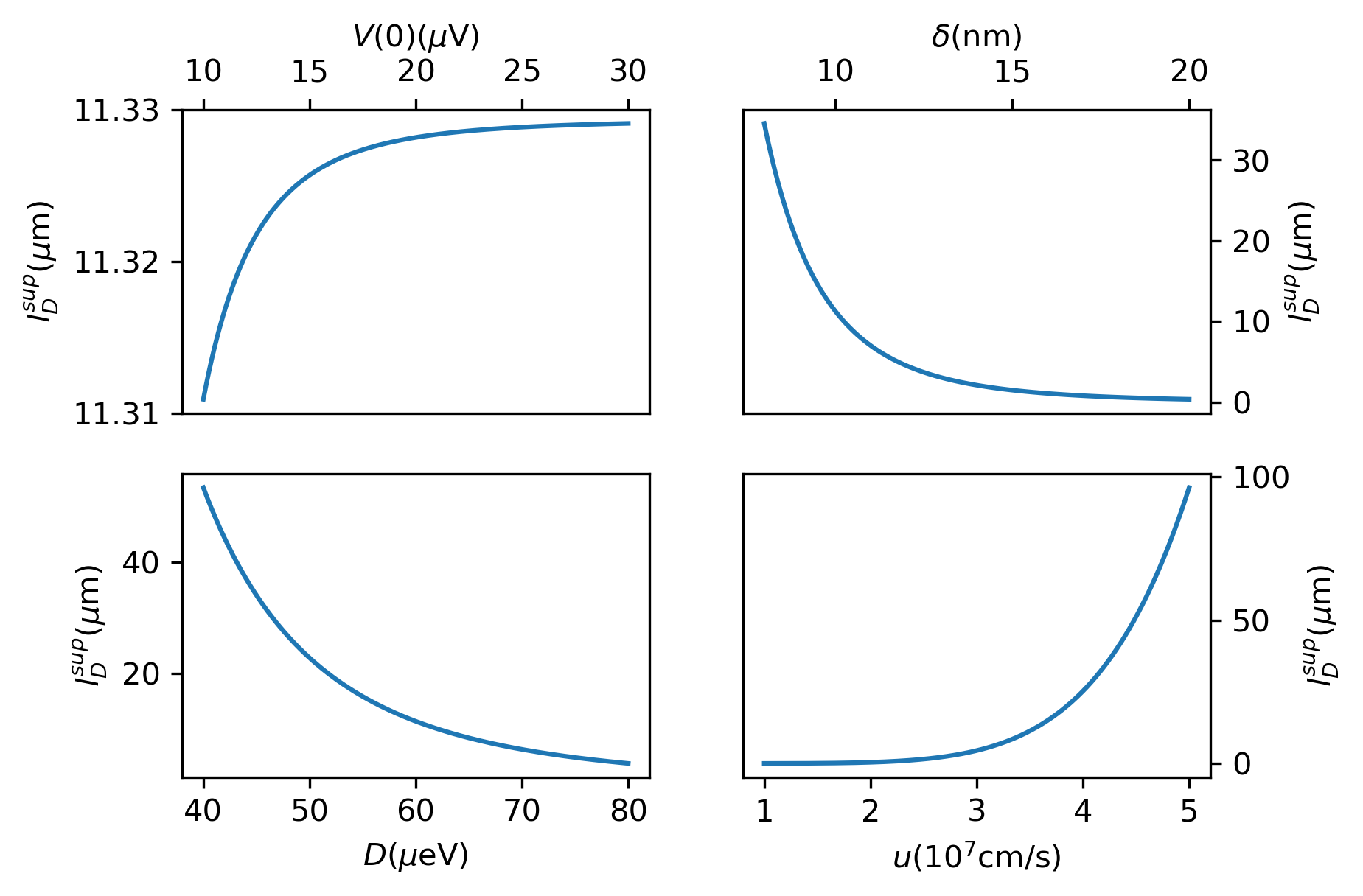}
\caption{\label{fig:maximum edge length} Maximum edge length with respect to initial applied voltage $V(0)$, edge channel width $\delta$, spin gap D and edge velocity $u$. For the length dependence on any of the parameters, values of other parameters are taken from the set $D=60\mu$eV, $\delta=100$\AA, $V(0)=20\mu$V, $u=3.5*10^7\text{cm/s}$ and $1/\beta=2\mu$eV.} 
\end{figure}

Eq.(\ref{eq:relate polarization to that of 5T}) does not consider the role of the equilibration process. In a sufficiently long domain wall, the decrease in the chemical potential difference will ultimately make it on the order of temperature and will preclude further nuclear spin polarization.
We can use Eq.~(\ref{eq:voltage width relation}) to estimate an upper bound on  $l_D$ that would lead to polarization of nuclear spins along the entire domain wall.  The chemical potential difference between the helical channels at the end of the strip $\mu\left(l_D\right)$ must be 
\begin{equation}
\mu\left(l_D\right)\geqslant \mu_D=\frac{2}{\beta} \tanh^{-1}{\left(\frac{D}{E_{\rm O}}\right)}
\end{equation}
This then gives the upper bound of $l_D$ as:
\begin{equation}\label{eq:length restriction}
l_D\leq\frac{\mu_D^{2-2/K}-\mu(0)^{2-2/K}}{\pi \hbar  C_0\left(1-D/E_{\rm O}\right)(2/K^2-2/K)}=l_D^{\text{sup}}.
\end{equation}
For fixed $l_D$, the chemical potential difference due to applied voltage $eV=eV(0)=\mu(0)$ must somewhat exceed $\mu_D$ to achieve saturation spin polarization for all nuclei in the domain wall. However, in the $g=3$ case, due to the large exponent, if to aim for a full nuclear spin polarization of the domain wall and a subsequent nuclear spin polarization-induced spin transition in the injector region, a more effective way can be to reduce the spin gap $D$ using electrostatic gates. For parameters of the system that we used to calculate the nuclear spin polarization plotted in Fig.\ref{fig:polarization-voltage_dependence} we estimate $l_D^{\text{sup}}\approx 10\mu\text{m}$. In Fig.~\ref{fig:maximum edge length} we plot the dependence of $l_D^{\text{sup}}$ on the applied voltage $eV(0)=\mu(0)$, the spin gap in the interior regions $D$, edge state velocity $u$ and short-range cut-off $\delta$ of the helical modes.
We observe that the maximum length is rather insensitive to $V(0)$ due to a large exponent value in Eq.(\ref{eq:length restriction}). However, the maximum length strongly depends on $\delta$ and $u$, because $\delta$ determines the number of nuclear spins contributing to the tunneling current and $u$ defines the propagation speed of the edge-like states. These results show that such a boundary shift mechanism can be effective in a device on a length scale of the order of $10\mu\text{m}$ when the parameters are properly tuned. This correlates with the experiments \cite{Wang2021}, where the large resistance of the domain walls, which in the Hall bar configuration at $\sigma_{xx}\ll \sigma_{xy}$ corresponds to a maximum of the domain wall conductance, was observed in domain walls with a length of $2$ to $7\mu\text{m}$. After passage of a large $1$nA current, which corresponds to $\sim 40\mu$V voltage, for several minutes, the resistance increased in a $7\mu\text{m}$ domain wall sample, which can be interpreted as the suppression of tunneling due to dynamic nuclear spin polarization. Furthermore, electrostatic gates required adjustment, indicating a displacement of the domain wall.

\begin{figure}
\centering
\includegraphics[width=7.2cm]{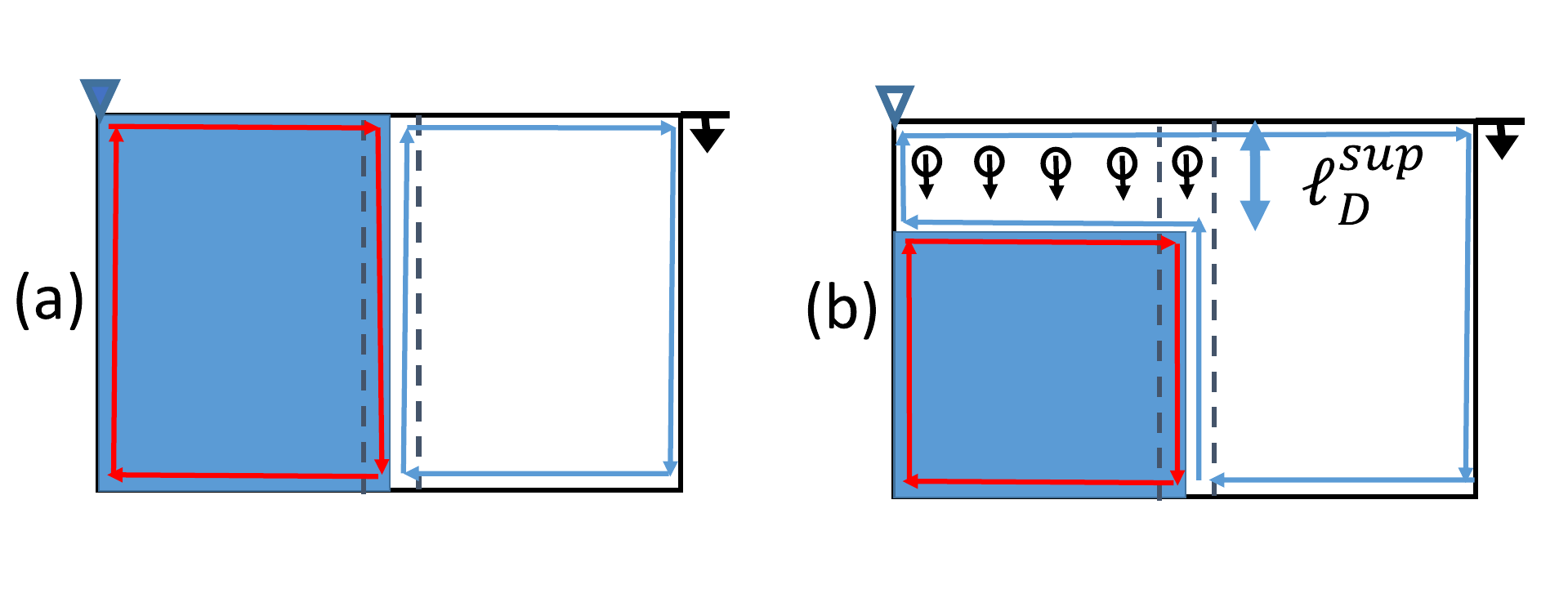}
\caption{\label{fig:when length greater then lD upper bound} Displacement of the domain wall in the case when $l_D>l_D^{\text{sup}}$, the blue shaded area is unpolarized phase, the unshaded area is the polarized phase. Red edge state is spin down and blue edge state is spin up (a) domain wall with initially unpolarized nuclei; (b) Current-induced nuclear polarization results in the displacement of the domain wall in a $l_D>l_D^{\text{sup}}$ wide strip. The polarized phase reaches the source contact, but will continue to evolve, see Fig.7. The lower segment of the domain wall with nuclear spin polarization insufficient to induce the domain wall shift remains unaltered.}
\end{figure}

We now analyze longer domain walls, $l_D>l_D^{\text{sup}}$. In this case, the domain wall nuclear spin polarization in the part of the device with coordinates along the domain wall $0< x \le l_D^{\text{sup}}$ will be sufficient to induce the Overhauser field capable of reversing the order of spin polarization of filled states, and therefore result in the boundary shift, as shown in Fig.~\ref{fig:when length greater then lD upper bound}. However, a decrease in the chemical potential difference due to equilibration effects can no longer lead to a sizable nuclear polarization that is sufficient to result in a change in the polarization  beyond that part of the domain wall. The path of the propagating Luttinger modes experiences a shift in the upper part of the device. This path goes clockwise or counter-clockwise as defined by the direction of the magnetic field. From the left side it goes horizontally, along the boundary of regions with polarized and unpolarized nuclear spins, where the order of filled and unfilled states has changed and returns to the original path along the part of the original domain wall and then along the true edge of the sample. In the course of a build-up in nuclear spin polarization, the regions, in which the order of filled levels reverse, do not extend over distances longer than $l_D^{\text{sup}}$ in the direction of domain walls arising during the build-up. However, once all regions, in which spin order is reversed, emerge, the length of the total path along the modified boundary between the  polarized and unpolarized regions may exceed the equilibration length. If this occurs, nearly all current will flow along the edge of the sample in the collector region, in which the spin order is not getting changed, as if the domain wall does not exist. The cause of this effective elimination of the domain wall is spin flips of tunneling particles that accompany the process of building up the current-induced nuclear spin polarization. 

We note that our solution of the kinetic equation Eq.~(\ref{eq:initial dM dt}) assumes that degrees of freedom for electron excitations are much faster than nuclei degrees of freedom. Still as nuclear spins get polarized they in turn affect electron modes, in particular, change the current path, as well as spin transitions. Then hyperfine interaction again lead to changes in the nuclear system. The equilibration length for the current-induced spin polarization sets an upper spatial scale at which non-equilibrium electron excitations propagating along the phase boundary affect the nuclear spins. However, changes in nuclear system incurred while this spatial scale is reached can affect electron excitations back. We will now see how this can result in oscillatory in space and time changes in coupled electron-nuclear system.   

\subsection{Oscillations of nuclear spin polarization in spin transitions near a source contact}
\begin{figure}
\centering
\includegraphics[width=0.45\textwidth]{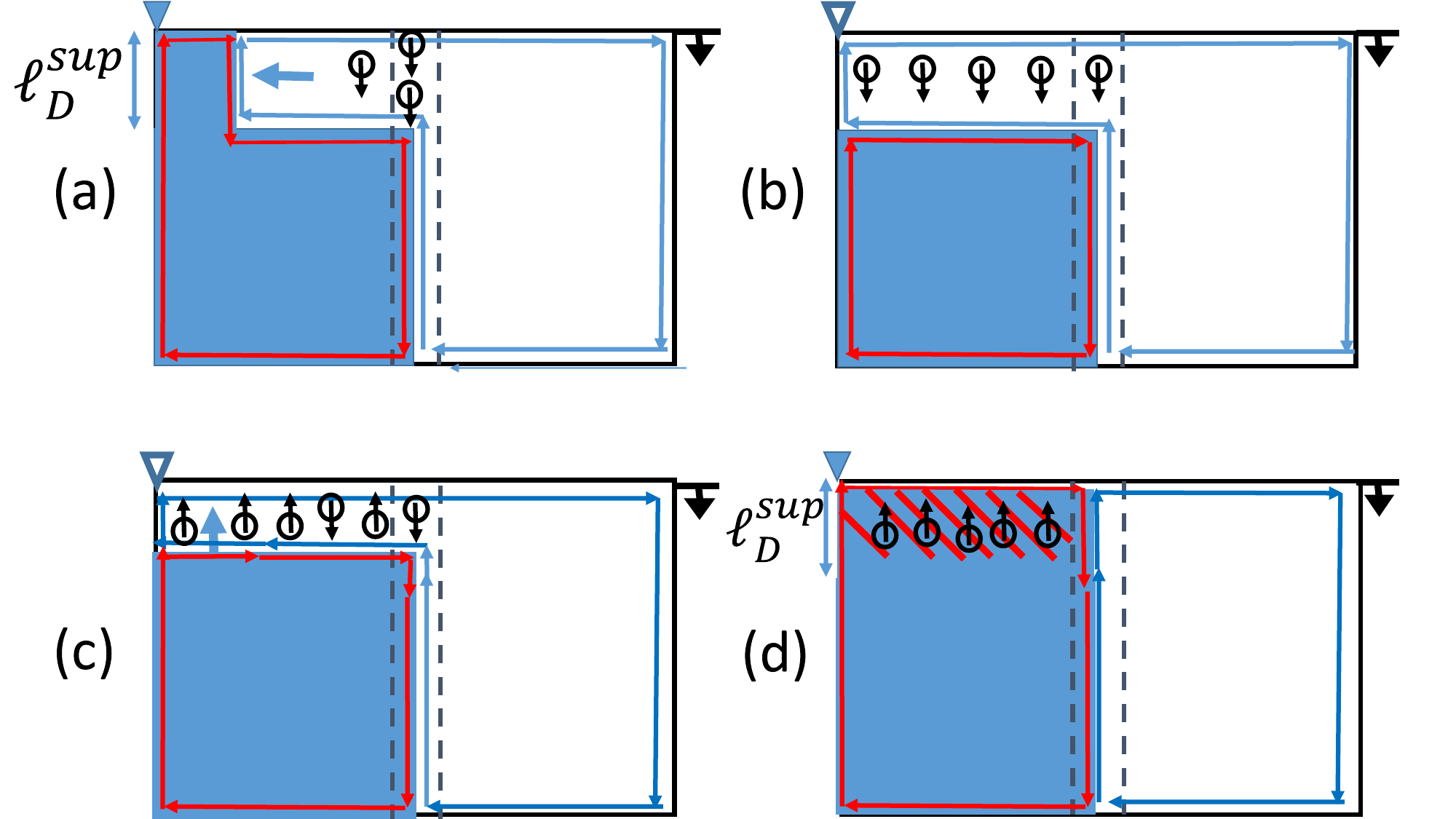}
\caption{\label{fOscillations} Oscillatory cycle of spin transitions near the source contact. Only edge channels of opposite spin polarization are shown at the boundaries of the sample and boundaries of the polarized and unpolarized regions; (a) Expansion of the polarized region in a strip with a $\ell_D^{sup}$ width towards towards the source contact to the unpolarized region (the source contact is shown as dark triangle in the left upper corner in the schematic drawing of the sample), due to current-induced nuclear spin polarization. Nuclear spins are shown as circles with arrows; (b) The area around the source contact is polarized and the injector region in a strip with a $\ell_D^{sup}$ width is polarized, the unpolarized region is no longer contacted by the source contact, spin pumping stops; (c) Due to boundary at $\ell_D^{sup}$ being near spin transition, with counter-propagating channels being an efficient channel of nuclear spin relaxation, nuclear spin relaxation leads to decrease of spin at the boundary, and the unpolarized region begins to expand towards the polarized injector, including the area near the source contact, as shown by the blue arrow ; (d) The source area is back in the unpolarized phase, and spin pumping resumes again. The dashed area in the left upper corner with a side length $\ell_D^{sup}$ is the area undergoing oscillatory back and forth spin transitions in time.}
\end{figure}

When the injector region undergoes spin order reversal due to nuclear spin polarization, this can also happen in the area of the source contact. Then, two effects occur. First, as soon as the nuclear spin polarization induces a phase change around the injector contact, the voltage drop occurs entirely within the part of the sample in a common FQHE polarization phase, and spin pumping  stops. The current path in the area with source and drain flows in the direction determined by magnetic field. The injected current from the contact flows from the source to the drain on the upper edge, no current flows on the lower edge due to grounded drain contact. Second, there is a new boundary between the polarized and unpolarized phases. Edge states counter-propagating at this boundary at $\ell_D^{sup}$ provide an efficient channel of nuclear spin relaxation, which defines nuclear polarization in accord with kinetic equation (\ref{eq:initial dM dt}). In the absence of spin pumping because of location of source and drain within a common spin FQHE phase, nuclear spin polarization near the boundary must evolve towards new steady state, more precisely towards equilibrium nuclear spin polarization. The nuclear spin relaxation on the injector contact side decreases the effective Overhauser field and must lead to a reverse spin transition. We now demonstrate that the area near the contact exhibits back and forth spin phase transitions. 

To be specific, let us assume that initially there was an unpolarized phase in the injector region and a polarized phase in the collector region. Due to domain wall displacement the area near the source contact eventually becomes polarized, Fig.7a,b. The nuclear spin polarization in the whole area near the contact is spin down, because the injection from the unpolarized region to the polarized region leading to the spin-flip current involves flip of electron excitation spin from down to up and the flip of nuclear spin from up to down. Once the source contact is located within the polarized region, spin pumping stops, and nuclear spin relaxation towards new steady state leads to transitions of nuclear spin down to nuclear spin up. That decreases the magnitude of the nuclear spin polarization, Fig. 7c. One of the most efficient channels of the nuclear spin relaxation is hyperfine interactions between nuclear spins and spins of edge modes near the injector contact. As nuclear spins flip from down to up, electrons tunnel from the polarized region with blue edge channel spin up to the unpolarized region with red edge channel with spin down. This is an example of non-equiliibrium nuclear spin resulting in tunneling between edge states when pumping has stopped. The relaxation of nuclear spin results in a lower Overhauser effective field and in a gradual reversal of the spin order in the source contact region back to the original unpolarized phase. 
Once this reversal occurs, Fig.7d, the potential difference between the source and the drain restores spin pumping again and induces tunneling across the emerging domain wall between the unpolarized injector region and the polarized collector region, and the spin-flip current in its helical channels. That in turn starts restoration of the nuclear spin polarization with spin down and results in the shift of the domain wall in the injector region. Thus, the injector region can undergo back and forth spin order reversal with time in the shaded square with dashed lines in Fig. 7d. 

Due to equilibration effects, the area where this back and forth phase change occurs is a strip with a side $l_D^{\text{sup}}$. We note that
the polarization of nuclear spins at $l_D^{\text{sup}}$ is precisely such that there is nearly a zero gap in composite fermion spectrum. Any few spin flips in conditions when the source contact is the polarized region start reversal back to the unpolarized phase, which proceeds till the contact is back on the unpolarized side. Thus both spin pumping and spin relaxation results in the domain wall motion, which then continues back and forth process. Using the upper boundary for the speed of the domain wall displacement $10^{-5}{\rm cm/s}$ calculated earlier, we estimate that the transition between two phases in the square containing the source contact with a side $l_D^{\text{sup}}=12\mu{\rm m}$ will take $2~{\rm minutes}$. The predicted oscillations can shed light on the data on the time dependent resistance observed in \cite{Zhong2018}.

\subsubsection{Role of the location of the source contact}
\begin{figure}
\centering 
\includegraphics[width=0.45\textwidth]{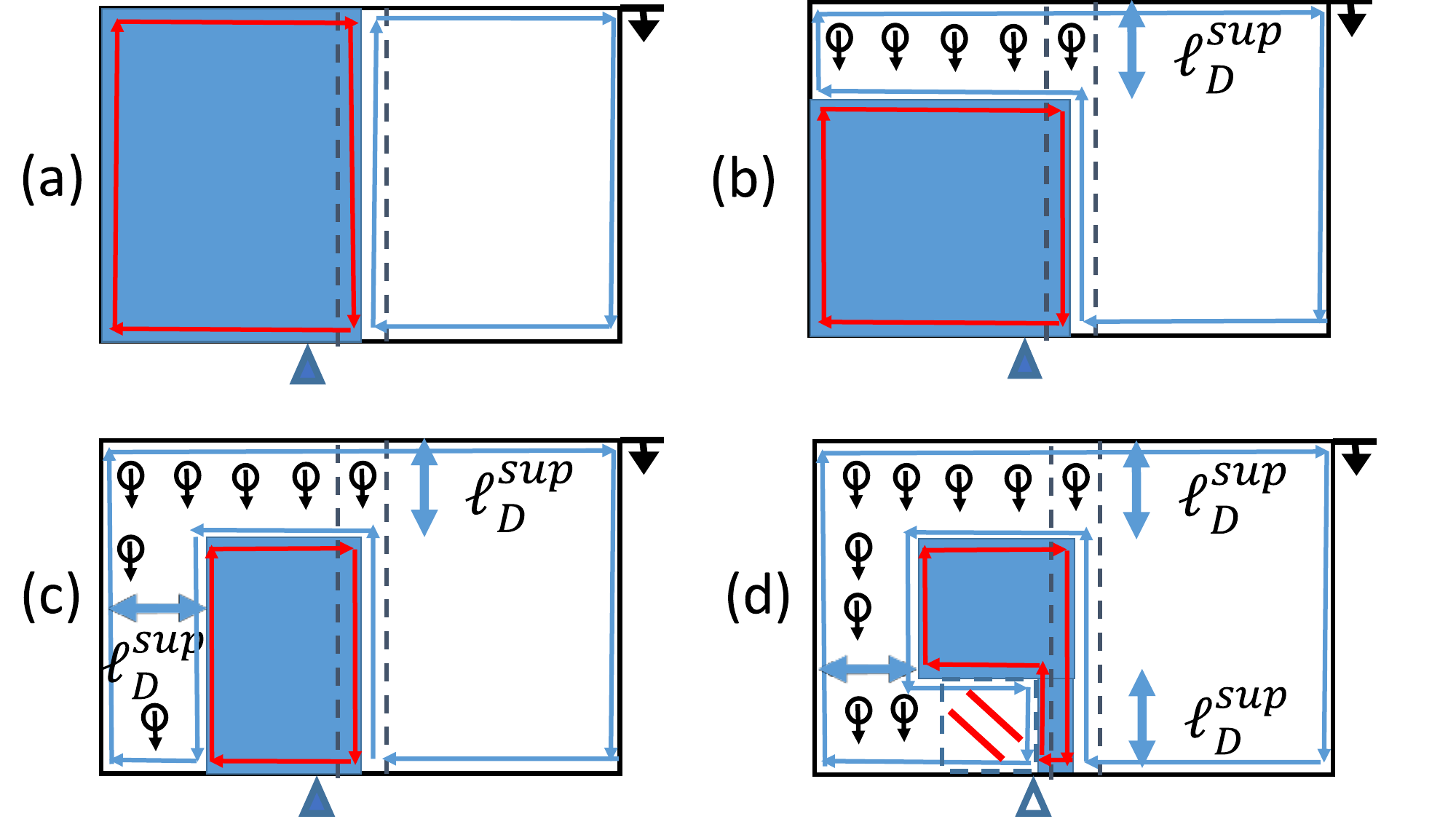}
\caption{\label{Location} Nuclear spin polarization and spin phase patterns for the source contact located on the lower edge of the sample near the domain wall between the polarized and unpolarized regions. (a) Domain wall with unpolarized nuclear spins at the initial stage. (b) Expansion of the polarized region towards source contact to the unpolarized region (shown in dark triangle in the left right corner) in a strip with a $\ell_D^{sup}$ width due to current-induced nuclear spin polarization. Due to the chirality of the charge motion in magnetic field this expansion proceeds first at the top of the sample. (c) The polarized phase expands towards the bottom edge in the left lower corner in a strip with a $\ell_D^{sup}$ width due to continued current-induced nuclear spin polarization (d) The polarized phase continues the expansion towards the source contact in the bottom edge of the sample. Once the source contact is located in the polarized region, the region of the dashed square with a side of length $\ell_D^{sup}$ becomes the stage area undergoing oscillatory back and forth spin transitions in time due to alternating spin relaxation and spin pumping.}
\end{figure}
We further consider how the location of the source contact in the device affects the nuclear spin polarization pattern. The contact is not necessarily located in the top left corner of the device, as in Fig.7 and can be located anywhere within the injector phase region. We are going to show now that in the magnetic field, which defines the direction of propagation of modes along the edges, placing the contact on the lower edge can lead to non-trivial pattern of nuclear spin polarization within the sample as in Fig.8. 

Suppose the source contact is close to the domain wall on the lower true edge of the sample in the polarized region at a distance $L > l_D^{\text{sup}}$ from the left sample edge, the drain contact is in the right corner of the sample in the polarized region, and the direction of magnetic field dictates clockwise direction for charge motion. Following the clockwise direction, the current mode reaches the domain wall and a part of the current flows along the domain wall, inducing nuclear spin polarization. Then, similarly to Fig.6a, the unpolarized mode gradually extends towards the upper left corner of the sample, forming a rectangle, the width of which  is limited by $l_D^{\text{sup}}$ due to the equilibration. Now that the contact is not in that upper left corner, when the polarized phase reaches the left edge of the sample, the domain wall between the polarized and unpolarized phase with a centerline at $x=l_D^{\text{sup}}$ continues its shift towards the injector side now located in the lower left corner of the sample at $x>l_D^{\text{sup}}$ and $y<0$, until it reaches the bottom sample edge. At that stage the domain wall between the polarized injector region and the unpolarized collector region has its centerline at a distance  $l_D^{\text{sup}}$ from the left edge of the sample. The domain wall now continues its shift to the right, towards the source contact. 

We now suppose first that the width of the Hall bar in $x-$ direction is larger than $2 l_D^{\text{sup}}$, as drawn in Fig.8. Once the region of the source contact becomes polarized, the unpolarized region only consists of the narrow strip to the right of the source and to the left of the centerline of the original domain wall $y=0$ and an area in the middle of the sample surrounded by rectangles with widths $l_D^{\text{sup}}$ that extend until the sample edges at the top, at the bottom and on the left. Because the area of the source contact at this stage has transitioned to the unpolarized phase from the polarized phase, spin pumping stops, and spin relaxation begins, Then the area near the contact with a size $l_D^{\text{sup}}$ starts undergoing oscillatory change of nuclear spin polarization and oscillatory back and forth FQHE spin transitions between the polarized and unpolarized phases. 

In the case that the width of the Hall bar in $x-$ direction is smaller or equal than $2 l_D^{\text{sup}}$, the process of the domain wall shift and expansion of the polarized phase towards the source contact results in transition to the polarized phase almost in the whole sample with the exception of the narrow strip to the right of the source and to the left of the centerline of the original domain wall $y=0$. Once the polarized region reaches the source contact, the near the contact with a size $l_D^{\text{sup}}$ becomes the stage of the oscillatory back and forth FQHE spin transitions between the polarized and unpolarized phases.   
 
\subsection{On the Overhauser effect on edge states}

We conclude this section with a discussion of the physical mechanism of the effect of the Overhauser nuclear spin magnetic field on the edge states. 
It is important to realize that while the spin flip current of the edge modes and a change in nuclear spin polarization are determined by the spin flip terms of Eq.(\ref{Hf1D}), the Overhauser effect on electrons and all correlated electron modes is defined via the Hamiltonian of the interacting bulk electrons Eq.(\ref{eq:theH}). The nuclear spin polarization acts on bulk electrons and modifies the spectrum of composite fermions, which in turn leads to modification of the edges. It is important that over longer periods of time used for spin pumping in the FQHE studies, the nuclear spin polarization changes both due to the displacement of the boundary and due to nuclear spin dipole-dipole interactions resulting in spin diffusion. 
We also note that the $I_z$ part of Eq.(\ref{Hf1D}) does not affect the tunneling current. However, it contributes to an Overhauser field acting on the edge modes. The $I_z$ term of Eq.(\ref{eq:theH}) can actually be removed by a gauge transformation \cite{Kivelson}. It then contributes to an extra phase term for bosons, which means a shift of the boson field. Still, to account for effects of all nuclei on tunneling current, it is important to consider modification of correlations via the spectrum of electrons in the whole of 2D liquid. An important example of the effect that would not emerge from the $I_z$ term of Eq.(\ref{Hf1D}) is the displacement of the domain wall, the displacement of the position of the level crossing within the domain wall, and the change of the current path.

\section{Role of Nuclei in Engineering Parafermions}\label{sec:fate of parafermion}

It has been suggested in Refs.~\cite{Alicea,Liang2019} that proximity-coupling of helical states that counter-propagate in a domain wall to an s-type superconductor can generate topological superconductivity and lead to localized parafermion zero modes on the border between the topological and an s-type superconductor or on the boundary of regions with prevailing superconducting gap and a prevailing normal gap. The normal gap is caused by an induced spin-flip tunneling. It has been proposed that coupling the system to a strong spin-orbit insulator~\cite{Alicea} can generate such a gap. However, for composite fermions arising from the ground Landau level, the spin-orbit interactions are negligible, as all matrix elements between ground Landau level states vanish; spin-orbit coupling emerges only as a result of coupling of the opposite spin zeroth and first Landau level electron states. In ~\cite{Liang2019} signatures of parafermion modes were obtained in the model with an in-plane magnetic field that induces spin-flip tunneling events. 

Remarkably, in the presence of electron spin to nuclear spin hyperfine spin-flip interactions,
a non-superconducting gap is generated even in the absence of an in-plane magnetic field. 
Electrons and therefore electron excitation modes are subject to the fluctuating Overhauser field of nuclear spins. 
Nuclear spin fluctuations are especially important if nuclear spins are only weakly polarized.
The average square of the effective nuclear spin field of the frozen nuclear fluctuation $\langle B^2\rangle$ \cite{Merkulov} 
is given by
\begin{equation}
    \langle B^2\rangle= \frac{16B^2_{\rm O}(I+1)}{3IN}.
\end{equation}
In the case under consideration, $N$ is the number of nuclei in the part of the domain wall, with a length of $l_t$, sandwiched between two proximity superconducting regions. This field defines a gap emerging instead of the crossing of spin levels, $\Delta_{\rm O}\propto \sqrt{\langle B^2\rangle}$, of non-superconducting origin. 
For this gap in the region of the domain wall with dimensions defined by $w_D=180$\AA, $\ell_z=120$\AA, and $l_t= 0.12\mu\text{m}$ we obtain $\Delta_{\rm O}=1\mu$eV. Therefore, in this case, for the magnitude of the proximity induced superconducting gap $\Delta_{sc}=1\mu$eV and characteristic temperatures used in the FQHE experiments $T=10$mK,  the configuration of the experiment discussed in \cite{Alicea,Liang2019} is promising for the observation of the parafermion modes.

Thus, nuclei are sufficient on their own; an additional transverse magnetic field can be used to tune the tunneling gap. As a position-dependent spin-flipping interaction, the frozen nuclear spin fluctuation is fully analogous to the spin-orbit interactions (the latter coupling, however, does not emerge in our system). Furthermore, one can also control the spin-flip tunneling by simply changing the value of the applied voltage/current through the system, taking it closer to or farther from saturation of the z-component of the nuclear spin polarization, which suppresses the tunneling process. The hyperfine interaction should also facilitate the proposal in~\cite{Universal_quantum}, which similarly requires the generation of tunneling non-superconducting gaps between the opposite spin quasiparticle modes.

\section{Discussion}\label{sec:discussion}

In this work, we have considered primarily the nuclear spin $1/2$. However, in general, nuclear spins can have different quantum numbers, such as $3/2$. For larger spin values, the renormalization group (RG) flow of the hyperfine interaction should behave similarly to that of a multi-channel Kondo problem. A model describing a multi-channel Kondo problem interacting with strongly correlated fractional quantum Hall (FQH) edge state electrons remains an open question to the best of the authors' knowledge. Therefore, we attempt to provide a qualitative discussion here. 

For a higher spin $\tilde{I}$ nuclear spin, we may approximately decompose it into $2\tilde{I}$ quantum number $1/2$ spins. This decomposition can then be effectively translated into a rescaling of $\rho_n$ by a factor of $2\tilde{I}$. As a result, the upper bound of $l_D$ will be smaller compared to the $1/2$ nuclear spin setting.

In Ref.~\cite{Adagideli}, a device utilizing oppositely polarized edge states of the two quantum anomalous Hall effect insulator electrodes was proposed to polarize nuclear spins in the central region between the two insulators, realizing a memory device approaching the Landauer limit. It follows from our consideration that the area of the electrodes close to the central region may experience the Overhauser field of nuclear spins. The dynamic nuclear spin polarization region can propagate further into the electrodes because of the dipole-dipole interactions and spin diffusion. If the induced effective Zeeman energy is strong enough to alter the ground state of the affected part of one of the anomalous Hall electrodes, a considerably larger part of the device (with asymmetric locations of the spin-up and spin-down TI phases), rather than just the central region, could experience the Overhauser effect and, therefore, could be used to store quantum information. 
Also, certain regions can potentially undergo oscillatory spin phase transitions, which might be of interest to utilize for applications.

\section{Conclusion}\label{sec:Conclusion}

In this work, we have studied the dynamics of nuclear spins at the boundary between polarized and unpolarized quantum Hall edge states and their roles in fractional quantum Hall effect spin transitions. We also discussed the effect of hyperfine interactions on the parafermion zero modes in a sought after hybrid superconductor-fractional quantum Hall effect system. Under typical experimental conditions, we have obtained nuclear spin polarization as a function of temperature, bias voltage, and relaxation due to dipole-dipole interactions. The non-equilibrium nuclear spin  polarization increases with the ratio of bias voltage to temperature. However, most importantly, 
both nuclear polarization and tunneling currents depend on electrostatic gate voltages that are used to control the filling factors and gaps between filled and unfilled spectral modes in the polarized and unpolarized fractional quantum Hall liquids. In experiments, this leads to a strong resistance dependence on the filling factors of the polarized and unpolarized quantum Hall liquids defined by the electrostatic gates.

We then evaluated the effects of the nuclear Overhauser field on the phases of the fractional quantum Hall liquids around the boundary and found that when the nuclear polarization is sufficiently strong the induced Overhauser field may switch the system between polarized and unpolarized phases. This leads to a mechanism for spreading the dynamical nuclear spin polarization via the displacement of the boundary between these two phases, potentially polarizing the majority of the nuclear spins in the sample. We have shown that the spread of nuclear spin polarization and the domain wall displacement are strongly asymmetric and occur primarily in the injector region. We also found that if the displacement of the domain wall results in the change of polarization state of the fractional quantum Hall liquid at the source contact, the area that includes the contact undergoes an oscillatory back and forth spin transitions between polarized and unpolarized states. The size of the area is defined by the equilibration length that we calculated using the Luttinger liquid model for helical edge-like modes. The developed quantitative theory of dynamic nuclear spin polarization in integer and fractional quantum Hall systems describes saturation of spin polarization of nuclei and ways of controlling the electric currents and magnetization in quantum Hall effect devices.

We show that owing to the interaction of electron excitations with nuclear spins, the parafermion zero modes in hybrid s-superconductor fractional quantum Hall structures can potentially arise at the boundaries of nuclear spin-induced and proximity-induced superconducting gaps. Finally, our results explain several aspects of the experimentally observed asymmetry in the spread of nuclear spin polarization in polarized and unpolarized regions and shed light on earlier unexplained experimental results on time-dependent resistance in \cite{Rokhinson_1,Zhong2018}. Detailed discussion of these experiments is beyond the scope of the present work and will be addressed elsewhere. 

\begin{acknowledgments}
We acknowledge helpful discussions with J.~I.~V\"{a}yrynen, G.~Li and V.~V.~ Ponomarenko, and are grateful to L.~P.~Rokhinson for useful discussions and sharing unpublished experimental data. 
\end{acknowledgments}

\appendix
\section{Evaluation of correlators}\label{ap:Evaluation of correlators}
Here we evaluate the following correlators that determine the tunneling current Eq.(\ref{eq:fqhe current}) and the polarization of nuclear spins Eq.(\ref{eq:fqhe steady polarization}):
\begin{equation}
\label{comm}
\mathfrak{Re}\int_0^\infty dt e^{i(-eV)t} \langle [\tilde{\Omega}(0,t), \tilde{\Omega}^\dagger(0,0)]\rangle
\end{equation}
and 
\begin{equation}
\label{ant}
\mathfrak{Re}\int_0^\infty dt e^{i(-eV)t}\langle \{\tilde{\Omega}(0,t), \tilde{\Omega}^\dagger(0,0)\}\rangle,
\end{equation} 
where we omitted $E_{\text{nz}}\ll eV$, as discussed in the main text. We set $\hbar=1$, and restore it in the equations of the main text. We use the time-ordered four-point correlator \cite{Giamarchi,Xiaogang}:
\begin{eqnarray}\label{eq:4 points time ordered correlator}
&&\langle \mathcal{T}\psi_R^\dagger(t,x)\psi_L(t,x)\psi_L^\dagger(0,0)\psi_R(0,0)\rangle\nonumber\\
=&&(\frac{1}{2\pi\delta})^2(\frac{\beta u}{\pi\delta})^{-2/K}[\sinh{(\frac{\pi}{\beta}\{\frac{x}{u}-[t-i\epsilon \text{sign}(t)]\})}\nonumber\\
&&\times\sinh{(\frac{\pi}{\beta}\{\frac{x}{u}+[t-i\epsilon \text{sign}(t)]\})}]^{-1/K},
\end{eqnarray}
where $\mathcal{T}$ is time-ordering operator.
We then evaluate the correlator with the commutator of $\Omega$'s Eq.~(\ref{comm}):
\begin{eqnarray}\label{eq:comutation}
&&\mathfrak{Re}\int_0^\infty dt\langle [\tilde{\Omega}(x,t), \tilde{\Omega}^\dagger(0,0)]\rangle=2\mathfrak{Re}\int_0^\infty dt i\sin(-eVt)\nonumber\\
&&\times\langle\mathcal{T}\psi_R^\dagger(t,x)\psi_L(t,x)\psi_L^\dagger(0,0)\psi_R(0,0)\rangle.
\end{eqnarray}
 The correlator with the anti-commutator of $\Omega$'s Eq.~(\ref{ant}) is given by
\begin{eqnarray}\label{eq:anti commutation}
&&\mathfrak{Re}\int_0^\infty dt\langle \{\tilde{\Omega}(x,t), \tilde{\Omega}^\dagger(0,0)\}\rangle=2\mathfrak{Re}\int_0^\infty dt\nonumber\\
&&\times\cos(eVt)\langle \mathcal{T} \psi_R^\dagger(t,x)\psi_L(t,x)\psi_L^\dagger(0,0)\psi_R(0,0)\rangle.
\end{eqnarray}
\begin{figure}[h]
    \centering
    \includegraphics[width=1\linewidth]{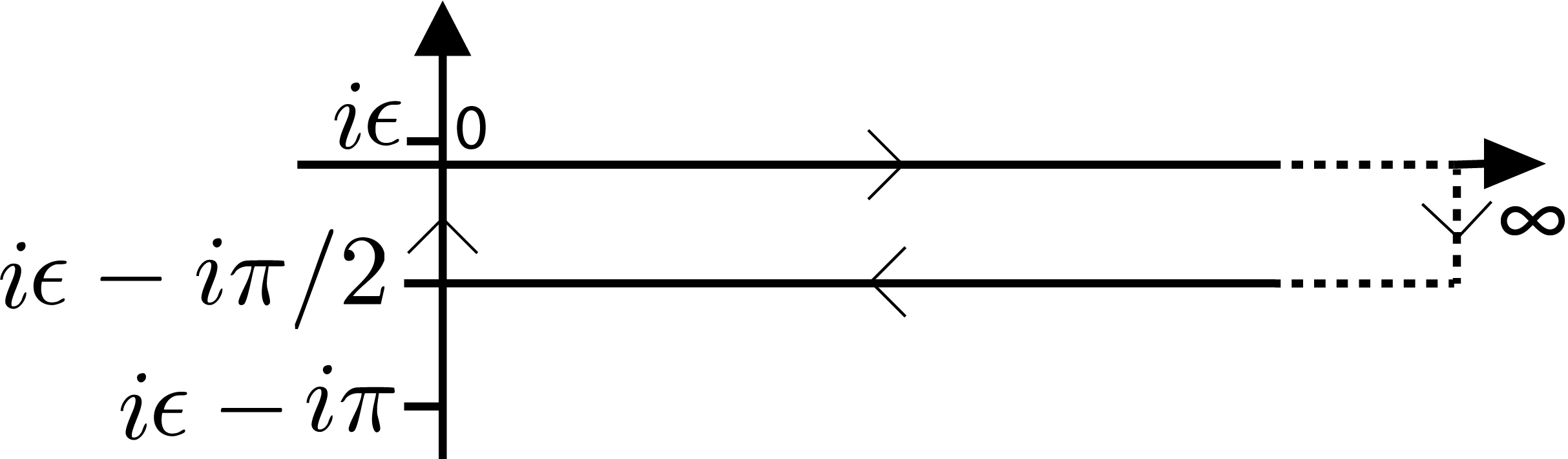}
    \caption{The arrowed path shows the contour used for integration}
    \label{fig:contour_integral}
\end{figure}
Setting $x=0$ and $t>0$, and noticing that both integrals have the same singularities at $t=i\epsilon+in\pi, n\in \mathbb{Z}$, we choose the integration contour as $0\rightarrow\infty\rightarrow\infty+i\epsilon-i\pi/2\rightarrow 0+i\epsilon-i\pi/2\rightarrow 0$, which is illustrated in Fig.~\ref{fig:contour_integral}. We temporarily ignore factors irrelevant to the integration and restore those in the final equations. Performing contour integration, we obtain: 
\begin{eqnarray}\label{eq:evaluate commutation}
&&\mathfrak{Re}\int_0^\infty dt'i\sin(-eVt')\sinh^{-2/K}{(\frac{\pi}{\beta}t'-i\epsilon)}\nonumber \\
&&=\mathfrak{Re}\int_0^\infty dt\frac{\beta}{\pi}(-i)\sin(\frac{eV\beta t}{2\pi})\sinh^{-2/K}(t-i\epsilon)\nonumber\\
&&=\mathfrak{Re}\int_0^{i\epsilon-i\frac{\pi}{2}}dt(-i)\frac{\beta}{2\pi}\sin(\frac{eV\beta t}{2\pi})\sinh^{-2/K}{(t-i\epsilon)}+\nonumber\\
&&\mathfrak{Re}\int_{i\epsilon-i\frac{\pi}{2}}^{\infty+i\epsilon-i\frac{\pi}{2}}dt(-i)\frac{\beta}{\pi}\sin(\frac{eV\beta t}{2\pi})\sinh^{-2/K}(t-i\epsilon),
\end{eqnarray}
where the first term on the RHS of the last equality can be rewritten as:
\begin{eqnarray}\label{eq:evaluate commutation first term}
&&\mathfrak{Re}\int_0^{\epsilon-\frac{\pi}{2}}dt\frac{\beta}{\pi}\sin(\frac{ieV\beta t}{2\pi})\sinh^{-2/K}{(it-i\epsilon)}\nonumber\\
=&&\mathfrak{Re}\int_0^{\epsilon-\frac{\pi}{2}}dt\frac{\beta}{\pi}i\sinh(\frac{eV\beta t}{2\pi})\frac{1}{i^{2/K}\sin^{2/K}{(t-\epsilon)}}
\end{eqnarray}
In the case of FQHE filling factor $\nu=K$, $1/K$ is an integer, and no branch cut is required. When there is no branch cut, the term becomes purely imaginary and results in a divergence, and is therefore dropped.
The second term can be rewritten as
\begin{eqnarray}
&&\mathfrak{Re}\int_0^{\infty}dt(-i)\frac{\beta}{\pi}\sin(\frac{eV\beta t}{2\pi}+i\frac{eV\beta}{2\pi}(\epsilon-\frac{\pi}{2}))\nonumber\\
&&\times\sinh^{-2/K}(t-i\frac{\pi}{2})=\nonumber\\
&&\mathfrak{Re}\int_0^{\infty}dt(-i)\frac{\beta}{\pi}\frac{(e^{\frac{ieV\beta t}{2\pi}}e^{\frac{eV\beta}{2}}-e^{-\frac{ieV\beta t}{2\pi}}e^{-\frac{eV\beta}{2}})}{2i}\nonumber\\
&&\times\frac{1}{i^{2/K}\cosh^{2/K}(t)}|_{\epsilon\to 0}=\nonumber\\
&&\mathfrak{Re}\int_0^\infty dt\frac{\beta}{\pi}\sinh(\frac{eV\beta}{2})\frac{\cos(\frac{eV\beta t}{2\pi})}{i^{2/K}\cosh^{2/K}(t)}=i^{2/K}4^{\frac{1}{K}-1}\frac{\beta}{\pi}\nonumber\\
&&\times\sinh(\frac{eV\beta}{2})B(\frac{1}{K}+\frac{ieV\beta}{2\pi},\frac{1}{K}-\frac{ieV\beta}{2\pi})
\end{eqnarray}
In the derivation of the latter integral, we applied formula (1) of the section $3.512$ in \cite{tableofint}. Restoring all factors, we have:
\begin{eqnarray}\label{eq:commute}
&&\mathfrak{Re}\int_0^\infty dt\langle [\tilde{\Omega}(0,t), \tilde{\Omega}^\dagger(0,0)]\rangle=(\frac{1}{2\pi\delta})^2(\frac{\beta u}{2\pi\delta})^{-2/K}\nonumber\\
&&\times\frac{\beta}{4\pi}\sinh(\frac{eV\beta}{2})B(\frac{1}{K}+\frac{ieV\beta}{2\pi},\frac{1}{K}-\frac{ieV\beta}{2\pi}).
\end{eqnarray}
Similarly, we can obtain:
\begin{eqnarray}\label{eq:anti-commute}
&&\mathfrak{Re}\int_0^\infty dt\langle \{\tilde{\Omega}(0,t), \tilde{\Omega}^\dagger(0,0)\}\rangle=(\frac{1}{2\pi\delta})^2(\frac{\beta u}{2\pi\delta})^{-2/K}\nonumber\\
&&\times\frac{\beta}{4\pi}\cosh(\frac{eV\beta}{2})B(\frac{1}{K}+\frac{ieV\beta}{2\pi},\frac{1}{K}-\frac{ieV\beta}{2\pi}).
\end{eqnarray}
\bibliography{apssamp}

\end{document}